\documentclass[aps,pre,twocolumn,groupedaddress,superscriptaddress]{revtex4-1}

\usepackage{amsmath}
\usepackage{amsfonts}
\usepackage{amssymb}
\usepackage{bm}
\usepackage{color}
\usepackage{graphicx}
\usepackage{mathtools}
\usepackage{chemfig}
\usepackage{stackrel}
\usepackage{cancel}
\usepackage[utf8]{inputenc}

\newcommand{\st}{{\rm st}}
\newcommand{\CVSQ}{{\rm CV}^2}

\begin{document}

\title{Hyperaccurate bounds in discrete-state Markovian systems}
\author{D. M. Busiello}
\email{daniel.busiello@epfl.ch}
\affiliation{Institute of Physics, \'Ecole Polytechnique F\'ed\'erale de Lausanne---EPFL, 1015 Lausanne, Switzerland}
\author{C. E. Fiore}
\affiliation{Universidade de São Paulo,
Instituto de Física,
Rua do Matão, 1371, 05508-090
São Paulo, SP, Brasil}

\begin{abstract}
Generalized empirical currents represent a vast class of thermodynamic observables of mesoscopic systems. Their fluctuations satisfy the thermodynamic uncertainty relations (TURs), as they can be bounded by the average entropy production. Here, we derive a general closed expression for the hyperaccurate current in discrete-state Markovian systems, i.e., the one with the least fluctuations, for both discrete- and continuous-time evolution. We show that its associated hyperaccurate bound is generally much tighter than the one given by the TURs, and might be crucial to providing a reliable estimation of the average entropy production. We also show that one-loop systems (rings) exhibit a hyperaccurate current only for finite times, highlighting the importance of short-time observations. Additionally, we derive two novel bounds for the efficiency of work-to-work converters, solely as a function of either the input or the output power. Finally, our theoretical results are employed to analyze a 6-state model network for kinesin, and a chemical system in a thermal gradient exhibiting a dissipation-driven selection of states.
\end{abstract}

\maketitle

\section{Introduction}

Stochastic thermodynamics \cite{de2013non,seifert2012stochastic,ciliberto2017experiments,tome2015stochastic,van2015ensemble} constitutes a unified theory to describe the nonequilibrium properties of mesoscopic systems, encompassing  molecular motors \cite{akasaki,tome2010}, colloidal particles \cite{seifert2012stochastic,ciliberto2017experiments}, chemical reaction networks \cite{rao2016nonequilibrium,esposito2020open,busiello2021dissipation}, and phase transitions \cite{PhysRevE.100.012104,PhysRevE.102.022101,PhysRevE.104.064123}. The nonequilibrium behavior of a system is typically characterized by a continuous dissipation of energy into the environment to eventually reach and maintain a stationary state. The energy supply to sustain this steady consumption might stem from the coupling to one \cite{proesmans2016linear,brandner2015thermodynamics,proesmans2015onsager} or multiple reservoirs \cite{fiorek,busiello2020coarse}, both considering fixed thermodynamic forces and time-dependent drivings \cite{raz2016mimicking,busiello2018similarities}.

The breaking of detailed balance, a positive total entropy production rate, the presence of steady probability currents, and a limited efficiency (in the case of thermal engines) are only a few possible fingerprints of a nonequilibrium picture. Some of these features are also intimately connected through the celebrated fluctuation theorems \cite{jarzynski1997nonequilibrium,seifert2012stochastic} and satisfy universal bounds known as thermodynamic uncertainty relations (TURs), generally dictating that the dissipation constraints current fluctuations out of equilibrium. TURs have attracted increasing attention in recent years, hinting at the fascinating perspective of estimating the entropy production by measuring stochastic currents \cite{li2019quantifying,van2020entropy,manikandan2020inferring}.

In its original formulation, the TUR relates fluctuations of any stochastic currents in steady state arbitrarily far from equilibrium to the average total entropy production rate, $\langle \Sigma \rangle$ \cite{barato2015thermodynamic}:
\begin{equation}
\frac{\sigma^2_J}{ \langle J \rangle^2} \geq \frac{2}{\langle \Sigma \rangle},
\label{TUR}
\end{equation}
where $\sigma_J^2$ and  $\langle J \rangle$ are the variance and mean of the current $J$ over the ensemble of stochastic trajectories, respectively, being such left side known as coefficient of variation squared ($\CVSQ$) of $J$. The TUR expressed in Eq.~\eqref{TUR} has been proven to hold both for Markovian discrete-state systems \cite{PhysRevE.96.020103} and for continuous-state systems following a Langevin dynamics \cite{PhysRevE.102.012120,PhysRevE.99.062126,PhysRevE.100.012134,PhysRevResearch.2.013060}. Subsequently, TURs have been extended to several cases, such as periodically-driven systems \cite{barato2019unifying} and discrete-time processes \cite{proesmans2017discrete,chiuchiu2018mapping}, in turn generating a wealth of novel bounds in stochastic thermodynamics \cite{gupta2020tighter,dechant2018multidimensional} and highlighting their connection with fluctuation theorems \cite{vroylandt2020isometric,hasegawa2019fluctuation}. Additionally, several TURs have been recently unified under a geometric interpretation \cite{falasco2020unifying}.

As stated before, besides the richness of its physical content, i.e. the minimum amount of dissipation required to have a current of a desired precision, TURs also play a leading role in estimating the average entropy production, $\langle \Sigma \rangle$, by inverting Eq.~\eqref{TUR}. Some works in this direction exploited the saturation of the bound in short-time experiments \cite{manikandan2020inferring}, even if the bottleneck of this inference problem relies on the ability to identify a current approaching the bound, so to provide a reliable estimate of $\langle \Sigma \rangle$. In \cite{busiello2019hyperaccurate}, a closed expression for the hyperaccurate current, i.e. the one minimizing the $\CVSQ$, is derived for a set of overdamped Langevin equations. This is clearly the best observable to bound the average entropy production rate using Eq.~\eqref{TUR}.

Here, we generalize the concept of hyperaccurate current to Markovian discrete-state systems. We derive a general closed expression for the hyperaccurate current in the case of both discrete (Markov chains) and continuous (Master Equation) time evolution. For systems with only one loop in the transition network (rings), we show that all currents have the same $\CVSQ$ in the long-time limit, while finite-time hyperaccurate currents can be defined. Conversely, in the presence of more than one loop, we derive the hyperaccurate current and its associated bound,
%\cancel{at and out of the steady-state}
both for finite times and in the long-time regime. The knowledge of the hyperaccurate current can also provide two novel bounds for the efficiency of general work-to-work converters, respectively as a function solely of the output or input work. We then illustrate our theory for two paradigmatic master equation systems, a six-state model for kinesin moving along a microtubule \cite{liepelt1,liepelt2}, and a chemical system in a thermal gradient exhibiting a dissipation-driven selection of states \cite{busiello2021dissipation}.

\section{Generalized empirical currents}

Consider a stochastic trajectory performed by a discrete-state system in the time interval $t \in [0,t_f]$. This is characterized by the set of visited states, $\{ x_i \}_{i = 0,\dots,N}$. The generalized empirical currents associated with this trajectory are defined as \cite{PhysRevE.96.020103}:
\begin{equation}
    J = \frac{1}{t_f} \sum_{ml} d_{ml} n_{ml},
    \label{current}
\end{equation}
where $n_{ml}$ is the number of jumps from the state $l$ to $m$ up to time $t_f$, and $d_{ml}$ the element $(ml)$ of an anti-symmetric matrix $\hat{d}$. The specific form of $\hat{d}$ determines the current. Clearly, $J$ is a trajectory-dependent quantity, since $n_{ml}$ depends on the set $\{ x_i \}_{i=0,\dots,N}$ as follows:
\begin{equation}
    n_{ml} = \sum_{k=0}^{N-1} \delta_{x_k,l} \delta_{x_{k+1},m},
    \label{nml} .
\end{equation} 
where  $\delta_{i,j}$ attempts to the Kronecker delta.

To evaluate the $\CVSQ$ for discrete-state systems, we compute average and variance of a generalized empirical current over all stochastic trajectories with the same duration $t_f$. From Eq.~\eqref{current} and by employing the anti-symmetric property of $\hat{d}$ together with the fact that it does not depend on the trajectory, the average current is given by
\begin{equation}
    \langle J \rangle  = \frac{1}{t_f} \sum_{m<l} d_{ml} j_{ml} ,
\end{equation}
where the sum now runs over all indices $m < l$, and $j_{ml}$ is the average current from the state $l$ to $m$:
\begin{equation}
    j_{ml} = \langle n_{ml} - n_{lm} \rangle.
\label{steady}
\end{equation}
Analogously, the variance of $J$ reads as follows:
\begin{equation}
    \sigma^2_J = \frac{1}{t_f^2} \sum_{mlm'l'} d_{ml} d_{m'l'} C_{mlm'l'},
\label{sigma1}
\end{equation}
where $C_{mlm'l'} = \langle n_{ml} n_{m'l'} \rangle - \langle n_{ml} \rangle \langle n_{m'l'} \rangle$. Using again the anti-symmetry of $d_{ml}$, we can restrict the summation in Eq.~\eqref{sigma1} over all indices $m < l$ and $m' < l'$, obtaining:
\begin{equation}
    \sigma^2_J = \frac{1}{t_f^2} \sum_{m<l, m'<l'} d_{ml} d_{m'l'} \mathcal{M}_{mlm'l'},
\end{equation}
with $\mathcal{M}_{mlm'l'} = C_{mlm'l'} + C_{lml'm'} - C_{mll'm'} - C_{lmm'l'}$. Later on we will determine the explicit form of $j_{ml}$ and $\sigma_J^2$ for Markov chains and master equation systems.

\section{Hyperaccurate currents\\ and bound}

The hyperaccurate current is determined by the matrix $\hat{d}$ that minimizes the $\CVSQ$, namely $\hat{d}^{(h)}$. Hence, for each element $d_{ij}$ we have to solve the following equation:
\begin{gather}
    \label{hyp1}
    \frac{\partial}{\partial d_{ij}} \frac{\sigma_J^2(t)}{\langle J(t) \rangle^2} \bigg|_{\hat{d} \to \hat{d}^{(h)}} = \\
    = \frac{2 \left( \langle J^{(h)} \rangle \sum_{\{ml\}} d^{(h)}_{ml} \mathcal{M}_{mlij} - \sigma^2_{J^{(h)}} j_{ij} \right)}{\langle J^{(h)} \rangle^3} = 0 \qquad \forall i,j , \nonumber
\end{gather}
where $\langle J^{(h)} \rangle$ and $\sigma_{J^{(h)}}^2$ correspond to the mean and variance of the hyperaccurate current, respectively, and $\{m,l\}$ is a short notation to denote that the sum is constrained to $m < l$ and $m' < l'$. Analogously, $\sigma_{J^{(h)}}^2$ is the variance of the hyperaccurate current. We can exploit the fact that $\CVSQ$ does not change when multiplying $\hat{d}^{(h)}$ by a constant. The solution of Eq.~\eqref{hyp1} is therefore defined up to an arbitrary factor. From now on, we shall fix this constant by setting $\sigma^2_{J^{(h)}}/\langle J^{(h)} \rangle = 1$. This procedure is analogous to the one employed in  \cite{busiello2019hyperaccurate}. Moreover, since both $\mathcal{M}_{mlij}$ and $j_{ij}$ diverge linearly with $t_f$ in the long-time limit, it is convenient to introduce the scaled quantities  $\Tilde{\mathcal{M}}_{mlm'l'}= \mathcal{M}_{mlm'l'}/t_f$ and $\Tilde{j}_{ml} = j_{ml}/t_f$ that stay finite when $t_f \rightarrow \infty$. With these choices, from Eq.~\eqref{hyp1}, the hyperaccurate coefficients $d_{ml}^{(h)}$ have to satisfy:
\begin{equation}
    \sum_{\{ml\}} d_{ml}^{(h)} \Tilde{\mathcal{M}}_{mlij} = \Tilde{j}_{ij} \qquad \forall i,j .
    \label{hyp2}
\end{equation}
Moreover, we arrive at the general expression for the \textit{hyperaccurate bound}, that is the minimum possible value of the $\CVSQ$ of any generalized empirical current:
\begin{equation}
    \mathcal{B}_h = \frac{1}{\langle J^{(h)} \rangle}.
    \label{hbound}
\end{equation}
It is worth mentioning that Eqs.~\eqref{hyp2} and \eqref{hbound} hold for any discrete-state system, whether it is described by a Markov chain or evolves according to a master equation, both at and out of the steady-state. In the next sections, we shall derive the statistics of the currents, namely mean and variance, for Markov chains and master equation systems, restricting ourselves to the relatively simple, yet quite general, case of stationary processes for simplicity.

%\section{Statistics of  currents for Markovian and master equation systems} 
\subsection{Statistics of currents for Markov chains}

%In this section, we  shall derive the hyperaccurate coefficients for generic markovian chains and master equation systems. 
%\subsection{Markovian chains}
Markov chains are characterized by a discrete-time evolution. Indeed, a transition between discrete states can only happen at a definite time interval, $\Delta t$. As a consequence, a trajectory of length $t_f$ will be necessarily constituted by $N = t_f/\Delta t$ transitions. Let $p_{i;t}$ the probability to be in the state $i$ at time $t$, the dynamics of a Markov chain is given by
\begin{equation}
    p_{i;t+\Delta t} = \sum_{j} A_{ij} p_{j;t},
\end{equation}
being indeed fully specified by the transition matrix $A_{ij} = p_{i;t+\Delta t|p_j;t}$. Hence, given a stochastic trajectory $\{ x_i \}_{i=0,\dots,N} := \bold{x}$, taking place in the time interval $t \in [0,t_f]$, its path probability reads:
\begin{equation}
    \mathcal{P}(\bold{x}) = p_{x_0;0} \prod_{i=1}^N A_{x_i x_{i-1}}.
    \label{path}
\end{equation}
This probability is properly normalized, i.e. $\sum_{\bold{x}} \mathcal{P}(\bold{x}) = 1$, with $\sum_{\bold{x}} := \sum_{x_0,x_1,\dots,x_N}$.

For stationary processes, the initial condition $p_{x_0;0}$ is equal to the steady-state probability distribution $p^{\rm st}_{x_0}$. Therefore, the average number of jumps from the state $l$ to the state $m$ is given by:
\begin{gather}
    \label{mean}
    \langle n_{ml} \rangle = \sum_{\bold{x}} \mathcal{P}(\bold{x}) \sum_{k=0}^{N-1} \delta_{x_k,l} \delta_{x_{k+1},m} = \\
    = \sum_{k=0}^{N-1} \sum_{\bold{x}} p^{\rm st}_{x_0} \prod_{i=1}^N A_{x_i x_{i-1}} \delta_{x_k,l} \delta_{x_{k+1},m} = N A_{ml} p^{\rm st}_l. \nonumber
\end{gather}
where we used the following properties of the transition matrix $A_{x_i x_{i-1}}$:
\begin{eqnarray}
\sum_{x_N,x_{N-1},\dots,x_{k+2}} \prod_{i=k+2}^N A_{x_i x_{i-1}} &=& 1, \qquad \textrm{and} \nonumber \\
\sum_{x_{k-1},x_{k-2},\dots,x_0} \prod_{i=1}^{k} A_{x_i x_{i-1}} p^{\rm st}_{x_0} &=& p^{\rm st}_{x_k}. \nonumber
\end{eqnarray}
Following a similar procedure, it is possible to compute the second moment, $\langle n_{ml}n_{m'l'}\rangle$, which is given by:
\begin{gather}
    \langle n_{ml}n_{m'l'}\rangle=\sum_{\bm x} P_{\bm x} \sum_{k=0}^{N-1} \delta_{x_{k},l} \delta_{x_{k+1},m} \sum_{k'=0}^{N-1} \delta_{x_{k'},l'} \delta_{x_{k'+1},m'} = \nonumber \\
    = \sum_{k=0}^{N-1} \sum_{k'=0}^{N-1} \sum_{\bold{x}} p^{\rm st}_{x_0} \prod_{i=1}^N A_{x_i x_{i-1}} \delta_{x_k,l} \delta_{x_{k+1},m} \delta_{x_{k'},l'} \delta_{x_{k'+1},m'},
    \nonumber
\end{gather}
where summations over $k$ and $k'$ in the equation above includes three kinds of terms: a first one including only trajectories in which
$k' > k$, a second one taking contributions from trajectories in which $k' < k$, and a third one accounting for the cases $k' = k$. By evaluating each term separately, we arrive at the following expression:
\begin{eqnarray}
\langle n_{ml}n_{m'l'}\rangle &=& \sum_{k=0}^{N-1} \sum_{k'=k+1}^{N-1} A_{m'l'} p_{l',k'\Delta t| m,(k+1)\Delta t }  A_{ml}~ p^{\st}_{l}    +\nonumber \\
&+& \sum_{k=0}^{N-1} \sum_{k'=0}^{k-1}   A_{ml} p_{l,k\Delta t| m',(k'+1)\Delta t }  A_{m'l'}~ p^{\st}_{l'}   +\nonumber \\
&+& \sum_{k=0}^{N-1} A_{ml}~ p^{\st}_{l} ~\delta_{ll'}~\delta_{mm'}.
\label{cov}
\end{eqnarray}
Hyperaccurate coefficients, and thus the  hyperaccurate bound, can be readily obtained substituting Eqs.~\eqref{mean} and \eqref{cov} into Eq.~\eqref{hyp2} for stationary processes described by a Markov chain.
%The equation determining the hyperaccurate coefficients, Eq.~\eqref{hyp2}, and the hyperaccurate bound, Eq.~\eqref{hbound}, can be readily obtained from Eqs.~\eqref{mean} and \eqref{cov} in the case of stationary processes described by a Markov chain.

\subsection{Statistic of currents for master equation systems}

To adapt the formalism developed in the previous subsection to master equation systems, it is sufficient to modify the transition matrix as follows:
\begin{equation}
A_{ij}= 
    \begin{cases}
       W_{ij} \Delta t &\qquad {\rm{if }} \;\;\; j \neq i\\
       1 - \sum_{j} W_{ji} \Delta t &\qquad \rm{otherwise},
    \end{cases}
\end{equation}
where $W_{ij}$ is now the transition rate from the state $j$ to the state $i$ and corresponds to the $(ij)$-th element of the transition rate matrix $\hat{W}$. As in the previous section, we consider time-independent transition rates to ensure that the system will eventually reach a unique stationary state. With this form of $A_{ij}$, the moments of the generalized currents can be obtained by performing the continuous-time limit, that is $\Delta t \to 0$. Hence, we obtain:
\begin{gather}
\label{meanME}
\langle n_{ml} \rangle = \int_0^{t_f} dt ~W_{ml} p_l^{\rm st} = t_f W_{ml} p_l^{\rm st}, \\
\langle n_{ml} n_{m'l'} \rangle = W_{ml} W_{m'l'} \int_0^{t_f} dt \int_0^{t} d\tau \Big( p_{l';t| m; \tau} p_l^{\rm st} + \nonumber \\ 
+ p_{l; t| m'; \tau} p_{l'}^{\rm st} \Big) + \delta_{mm'} \delta_{ll'} t_f W_{ml} p_l^{\rm st}
%\CEF{\langle n_{ml} n_{m'l'} \rangle = W_{ml} W_{m'l'} \int_0^{t_f} dt \Big(  \int_t^{t_f} d\tau  p_{l';t| m; \tau} p_l^{\rm st} }+ \nonumber \\ 
%\CEF{+  \int_0^{t} d\tau p_{l; t| m'; \tau} p_{l'}^{\rm st} \Big) + \delta_{mm'} \delta_{ll'} t_f W_{ml} p_l^{\rm st}.}
\label{sdevME}
\end{gather}
It is worth pointing out that the number of jumps occurring in a single trajectory is not fixed a-priori by its duration for master equation systems.

As outlined above, Eqs.~\eqref{meanME} and \eqref{sdevME} determine the hyperaccurate current and its associated bound. However, it is instructive to derive the equation for hyperaccurate coefficients explicitly in the long-time limit, i.e. $t_f \to +\infty$. Noting that:
\begin{equation}
    \langle n_{ml} \rangle = W_{ml} \frac{2}{t_f} \int_0^{t_f} dt \int_0^t d\tau ~p_l^{\rm st} ,
\end{equation}
and following the same procedure outlined in \cite{busiello2019hyperaccurate}, we arrive at the following expression for $\tilde{C}_{mlm'l'}$:
\begin{gather}
    \label{JtildeLTL}
    \tilde{C}_{mlm'l'} = W_{ml} W_{m'l'} \bigg( p^{\rm st}_{l'} \int_0^{+\infty} dt \left( p_{l; t| m'; 0} - p^{\rm st}_l \right) + \nonumber\\
    + ~p^{\rm st}_{l} \int_0^{+\infty} dt \left( p_{l'; t| m; 0} - p^{\rm st}_{l'} \right) \bigg) + \delta_{mm'} \delta_{ll'} W_{ml} p^{\rm st}_l.
    \label{CtildeLTL}
\end{gather}
As expected, there are no divergences in the long-time limit and the propagators only depend on time differences since we are considering stationary processes.
Since the expression for $\tilde{C}_{mlm'l'}$  enters into the definition of $\tilde{\mathcal{M}}_{mlij}$, we can write the equation for the hyperaccurate coefficients as follows:
\begin{equation}
    \sum_{\{ml\}} d_{ml}^{(h)} \tilde{\mathcal{M}}_{mlij} = J^{\rm st}_{ij}, \qquad \forall i,j
    \label{hyp3}
\end{equation}
where $J^{\rm st}_{ij} = W_{ij} p^{\rm st}_j - W_{ji} p^{\rm st}_i$ are the steady-state probability currents obtained from the master equation.

A useful way to handle Eq.~\eqref{CtildeLTL} is to expand all probabilities in terms of the eigenvalues and eigenvectors of transition matrix $\hat{W}$, i.e.
\begin{equation}
p_{l; t| m; 0} = p_l^{\rm st} + \sum_{i=2}^M v_l^{(i)} a_i^{(m)} e^{\lambda_i t},
\end{equation}
where $M$ is the total number of accessible states in the system, $v_l^{(i)}$ is the $l$-th component of the $i$-th eigenvector, and $\lambda_i$ its associated eigenvalue. Note that the eigenvalues are enumerated in descending order so that $\lambda_1 = 0 > \lambda_2 > \dots > \lambda_M$. Here, the initial conditions, i.e. the fact the system is in the state $m$ at time $0$, are encoded in the coefficients $a_i^{(m)}$ satisfying the following equations:
\begin{equation}
    p_l^{\rm st} + \sum_{i=2}^M v_l^{(i)} a_i^{(m)} = \delta_{l,m}, \qquad \forall l = 1,\dots,M \nonumber
\end{equation}
Hence, $\tilde{C}_{mlij}$ takes the following form:
\begin{gather}
    \tilde{C}_{mlm'l'} = W_{ml} W_{m'l'} \sum_{i\geq 2} \left( p^{\st}_{l'} \frac{v_l^{(i)} a_i^{(m')}}{\lambda_i} + p^{\st}_{l} \frac{v_{l'}^{(i)} a_i^{(m)}}{\lambda_i} \right) + \nonumber\\+\delta_{mm'} \delta_{ll'} W_{ml} p^{\rm st}_l.
    \label{Ceig}
\end{gather}
In all the numerical examples presented below, we implement Eq.~\eqref{Ceig} to estimate $d_{ml}^{(h)}$ through Eq.~\eqref{hyp3}.

\subsection{Finite-time hyperaccurate currents for rings}

\begin{figure}[t]
    \centering
    \includegraphics[width=\columnwidth]{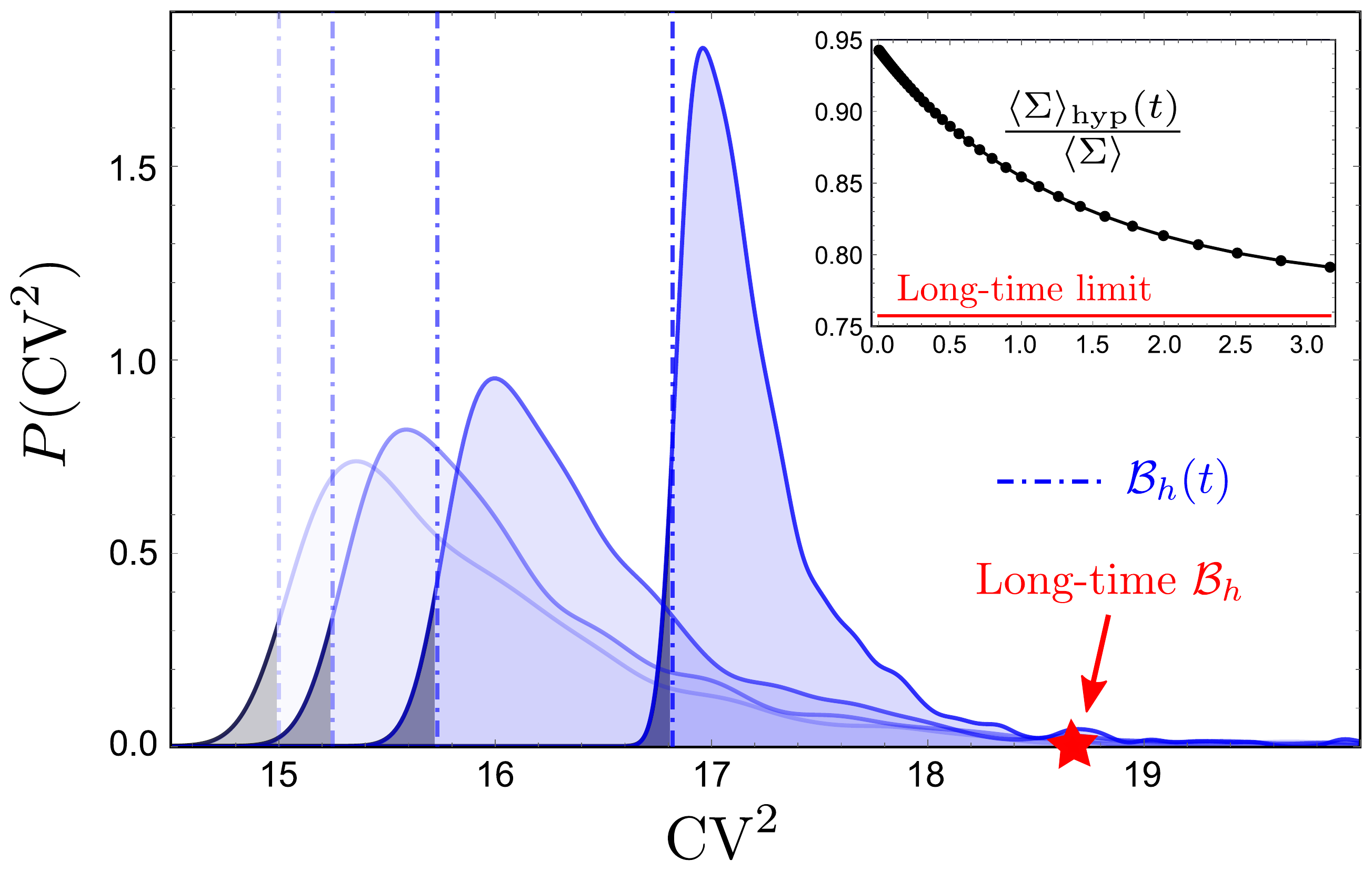}
    \caption{Finite-time hyperaccurate bound for a 4-state ring. \textit{Main}: The probability distribution of $\CVSQ$, $P(\CVSQ)$, is shown for increasing time, $t$, with decreasing opacity, and is compared to the finite-time hyperaccurate bound, $\mathcal{B}_h(t)$ (dot-dashed lines). The tails of $P(\CVSQ)$ that cross the line (gray areas) are consequences of the discretization employed in the histogram and do not correspond to any $\CVSQ$. Currents have been generated by perturbing the hyperaccurate solution. \textit{Inset}: The entropy production estimated from $\mathcal{B}_h(t)$, $\langle \Sigma \rangle_{\rm hyp}$, defined in Eq.~\eqref{EpEST}, approaches the actual value, $\langle \Sigma \rangle$, for short times. Here, their ratio is shown.}
    \label{figTIME}
\end{figure}

As a starting point, consider a discrete-state system constituted by one single loop (a ring)  in the transition network. At stationarity, all edges not belonging to the loop satisfy the detailed balance, and hence such a system can only support a unique nonzero steady current, $J^{\rm st}$.
%Thus, in the long-time limit, such a system can only support a unique steady nonzero current maintaining the nonequilibrium stationary state,
%Hence, for $t_f \to +\infty$, 
In the long-time limit, all possible currents of the system have to be proportional to $J^{\rm st}$, so that their $\CVSQ$ is equal to the one of $J^{\rm st}$, $\CVSQ_{\rm st}$:
\begin{equation}
    \CVSQ = \frac{\alpha^2 \sigma^2_{J^{\rm st}}}{\alpha^2 \langle J_{\rm st} \rangle} = \CVSQ_{\rm st} \nonumber
\end{equation}
where $\alpha$ is the proportionality factor.
%This means that, in the long-time regime, there is no hyperaccurate current for one-loop systems, since all currents have the same $\CVSQ$, irrespective of the number of states. An alternative proof of the absence of hyperaccurate currents for generic one-loop systems can be derived by inspecting the properties of the tilted matrix employed to evaluate the $\CVSQ$ within the large deviation formalism. However, we point out that this result holds only in the long-time limit. 
However, due to the transient regime of the propagators, it is possible to define a finite-time hyperaccurate current for any one-loop system, even in the presence of stationary processes. As illustrated in Fig.~\ref{figTIME} for a simple 4-state ring, the hyperaccurate bound changes over time (dot-dashed vertical lines with increasing opacity). As time increases, the probability distribution function of the $\CVSQ$ becomes narrower and narrower, eventually becoming a Dirac $\delta$ centered at $\mathcal{B}_h$ (red star) when $t \to \infty$, since all currents have the same $\CVSQ$ in the long-time limit. In the presented example, we have $2/\langle \Sigma \rangle < \mathcal{B}_h(t) \leq \mathcal{B}_h$, where the first inequality comes from the definition of the hyperaccurate bound. Hence, by inverting the finite-time hyperaccurate bound, one can get an improved estimation of the average entropy production at stationarity, namely $\langle \Sigma \rangle_{\rm hyp}(t)$. Indeed, we have:
\begin{equation}
    \langle \Sigma \rangle > \langle \Sigma \rangle_{\rm hyp}(t) = \frac{2}{\mathcal{B}_h(t)} \geq \frac{2}{\mathcal{B}_h}.
    \label{EpEST}
\end{equation}
In the inset of Fig.~\ref{figTIME}, we illustrate that the short-time estimation of the entropy production gets closer to the actual value of $\langle \Sigma \rangle$ than in the long-time limit. This finding has been found to be robust across all one-loop systems we numerically studied, suggesting that finite-time observations might be beneficial to estimate the dissipation of discrete-state Markovian systems, in line with a recent result obtained for overdamped systems \cite{manikandan2020inferring}. A proof of the generality of this property, along with the subsequent design of a feasible experimental procedure to take advantage of it, might be a fascinating topic for future works.

%currents different from $J^{\rm ss}$ for finite $t_f$, even when considering stationary processes, as a consequence of the transient regime of propagators. 
%Indeed, only in the long-time limit we can straightforwardly apply the large deviation argument presented above. 
%Thus, we employ our method to show how the finite-time hyperaccurate bound and the distribution %of currents change as $t_f$ increases. These results are reported in Fig.~1.

%Such findings are reported in Fig.~1.}

\section{Hyperaccurate efficiency bounds}

Hyperaccurate currents set the tightest possible bound to the entropy production rate of stochastic systems. As a consequence, they also provide us with general bounds on the efficiency of work-to-work converters that do not require a specific knowledge of system features,   being instead directly linked to the hyperaccurate bound $\mathcal{B}_h$. 

Work-to-work converters are a broad class of molecular engines operating at the nanoscale at a constant temperature (isothermal). They convert a given form of \textit{input work}, e.g., chemical, into a different form of \textit{output work}, e.g. mechanical, with a limited efficiency \cite{gupta2017stochastic}. They substantially differ from heat engines, as working substances do not undergo cyclic transformation between two different temperatures. Cellular transporters \cite{liepelt1,liepelt2,altaner2015fluctuating}, catalytic enzymes \cite{ma2016enzyme,busiello2020coarse}, Hsp70 chaperones \cite{de2014hsp70} are a few prominent examples of these machines in biology. 

The most relevant quantities in this scenario are the input, $\mathcal{W}_{\rm in}$, and the output power, $\mathcal{W}_{\rm out}$. Here, we derive two general upper bounds for the efficiency, each one depending on either $\mathcal{W}_{\rm in}$ or $\mathcal{W}_{\rm out}$. We start with the expression the steady-state average entropy production:
\begin{equation}
 \langle \Sigma \rangle = \sum_{(m,n)}\,(W_{mn}p^{\rm st}_n-W_{nm}p^{\rm st}_m)\ln\frac{W_{mn}}{W_{nm}},
\label{eq20}
\end{equation}
which can be rewritten in the usual bilinear form $\langle \Sigma \rangle = \sum_e J_e F_e$, where $F_e$ and $J_e$ are  thermodynamic forces and fluxes, respectively, and the sum runs over all fundamental cycles \cite{schnakenberg1976network}. Clearly, $J_e$ is in general a linear combination of some microscopic stationary fluxes, $J^{\rm st}_{mn}$. Analogously, $F_e$ will correspond to a combination of some microscopic forces, $F_{mn} = \log(W_{mn}/W_{nm})$.

To define an operating work-to-work converter, we consider the presence of a load and a drive force, respectively $F_l$ and $F_d$, with their corresponding fluxes, $J_d$ and $J_l$. Hence, we have:
\begin{equation}
    \langle \Sigma \rangle = J_d F_d + J_l F_l = \frac{1}{T} \left( \mathcal{W}_{\rm in} + \mathcal{W}_{\rm out} \right),
    \label{epp}
\end{equation}
where the right-hand side of this equation corresponds to the first law of thermodynamics with no variations of internal energy. To operate as an engine, one necessarily requires that the dissipation provided by the driving force, $\mathcal{W}_{\rm in} \geq 0$, generates a work that counteracts the external load, i.e. $\mathcal{W}_{\rm out} \leq 0$. Hence, the efficiency can be defined as follows:
\begin{equation}
    \eta = - \frac{\mathcal{W}_{\rm out}}{\mathcal{W}_{\rm in}} \in [0,1].
    \label{eff}
\end{equation}
By combining Eqs.~\eqref{epp} and \eqref{eff} together with the fact that $\mathcal{B}_h \geq 2/\langle \Sigma \rangle$ by construction, we obtain:
%one obtains the following bound relating $B_h$
%and $W_{out}$:
\begin{equation}
    \eta \leq \frac{B_h |W_{out}|}{2+B_h |W_{out}|} := \eta_{\rm b}^{\rm out}.
    \label{bound}
\end{equation}
Analogously, a second bound involving $\mathcal{W}_{\rm in}$ is derived:
\begin{equation}
    \eta\le 1-\frac{2}{B_h W_{in}} := \eta_{\rm b}^{\rm in}.
    \label{bound2}
\end{equation}
Notice that both these bounds do not require specific knowledge of system features, and they have to hold simultaneously at steady state. We stress the fact that $\eta_{\rm b}^{\rm out}$ requires, in principle, the measurement of the output power, whereas $\eta_{\rm b}^{\rm in}$ is solely based on the a-priori knowledge of the input work, e.g., the available chemical energy from ATP \cite{altaner2015fluctuating,de2014hsp70}. Moreover, it is possible to show that $\eta_b^{\rm out} \leq \eta_b^{\rm in}$, meaning that the knowledge of the output power provides a tighter bound to the efficiency. This finding also agrees with the naive expectation that  $W_{\rm out}$ is more informative than $W_{\rm in}$ to predict the efficiency of a work-to-work converter.

\section{Applications}
\subsection{Hyperaccurate current and efficiency in a model network for kinesin}

Kinesin is a molecular motor playing a fundamental role in biological processes, including mitosis, meiosis, and the transport of cellular cargo \cite{vale2000way,altaner2015fluctuating}. It consists of two amino acid chains forming a coiled coil with two motor heads on one end that are able to bind to microtubules. The other end of the dimer binds to cellular organelles. Kinesin performs processive walks on microtubules by subsequent binding and unbinding of the two heads. The hydrolysis of one ATP (adenosine triphosphate) into an ADP (adenosine diphosphate) and an inorganic phosphate (P) in the catalytic site placed in the motor head drives conformational changes that make the walk possible \cite{liepelt1,liepelt2}. It constitutes a remarkable example of a work-to-work converter since it transduces chemical energy into mechanical work.

\begin{figure}[t]
\includegraphics[width=0.98\columnwidth]{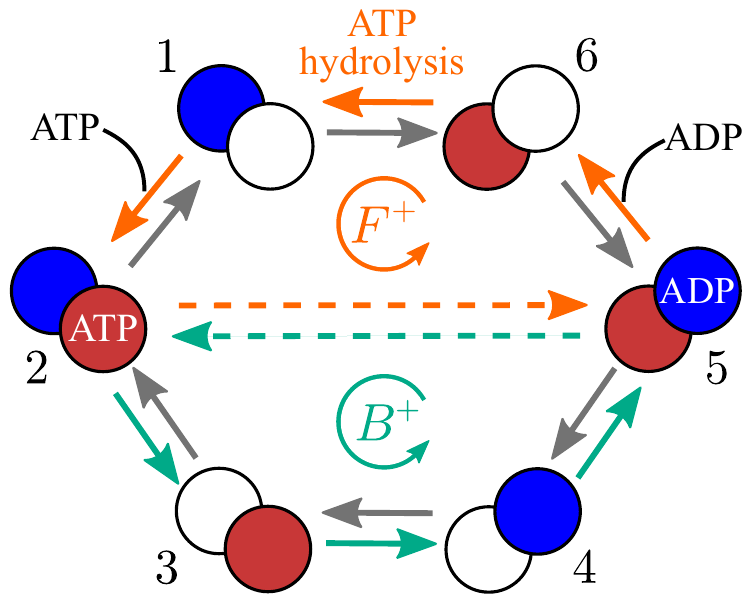}
\caption{Sketch of the 6-state kinesin model \cite{liepelt1,liepelt2}. $F^+$ and $B^+$ denote the forward and backward cycles, respectively. ATP states are indicated in red, while ADP states are in blue. The dashed arrows indicates the mechanical step, taken forward from 2 to 5, backward viceversa.}
\label{fig1}
\end{figure}

Here, we calculate both the hyperaccurate current and the efficiency bounds for kinesin, by applying the developed framework to the six-state transition network introduced in \cite{liepelt1,liepelt2}. Let us start with a brief introduction to the model. Each state is determined by the chemical composition of the two motor heads, e.g., ATP, ADP, or empty. Since we aim at describing processive motion, we ignore states in which both heads have the same composition \cite{liepelt1}. The network of all possible transitions is sketched in Fig.~\ref{fig1}. The system moves from the state 1 to 2 and from 4 to 5 via ATP binding; ADP binding drives the transition from the state 6 to 5 and from 3 to 2; the transition from the state 6 to 1 and from 3 to 4 are associated with ATP hydrolysis. Moreover, the dashed arrows identify the forward (from 2 to 5) and backward (from 5 to 2) mechanical steps.

There are six possible cycles in this network. $F^+ = 1 \to 2 \to 5 \to 6 \to 1$, indicated in Fig.~\ref{fig1}, encompasses the ATP hydrolysis and the subsequent forward step. Conversely, $B^+ = 2 \to 3 \to 4 \to 5 \to 2$ (see Fig.~\ref{fig1}) converts the energy from ATP hydrolysis into a backward step. Additionally, the system can also hydrolyze two ATP molecules, while performing no steps, following the purely dissipative cycle $D^+ = 1 \to 2 \to \dots \to 6 \to 1$, not reported in figure. Clearly, also the opposite cycles involving ADP synthesis can be performed, namely $F^-, B^-$, and $D^-$. The net processive walk is given by a competition between forward and backward cycles.

The dynamics of this system is controlled by two independent parameters, the dimensionless load force and $f=L{\tilde F}/k_{B}T$ (${\tilde F}$ and $L$ being the load force and the step size, respectively), and the chemical energy available from ATP hydrolysis,
\begin{equation}
    \Delta \mu = k_B T \ln \frac{K_{\rm eq} [{\rm ATP}]}{[{\rm ADP}] [{\rm P}]},
    \label{Dmu}
\end{equation}
in dilute conditions. The dynamics of kinesin can be by a master equation in which
the transition rate $W_{ij}$ from the state $j$ to $i$ is given by:
\begin{equation}
    W_{ij} = \kappa_{ij} \mathcal{I}_{ij}([X]) \Phi_{ij}(f)
\end{equation}
where $\kappa_{ij}$ is a constant, $\mathcal{I}_{ij}([X]) = [X]$ only if the reaction involves binding of $X$, otherwise it is equal to $1$, and $\Phi(f)$ takes the following form:
\begin{eqnarray}
    \Phi_{25}(f) &=& e^{-\theta f} \nonumber \\
    \Phi_{52}(f) &=& e^{(1-\theta) f} \\
    \Phi_{ij}(f) &=& \frac{2}{1+e^{\chi_{ij} f}} \qquad \forall (i,j) \neq (2,5) \nonumber
\end{eqnarray}
with $\theta$ and $\chi_{ij}$ additional constant factors. This choice of the transition rates agrees with the experimental observations, and also satisfies the energetic balance for each cycles \cite{liepelt1}. This condition states that detailed balance holds if no energy is available from ATP, otherwise chemical energy is converted into mechanical motion. For example, by inspecting the cycle $F^+$, we have the following energetic balance:
\begin{equation}
    k_B T \sum_{(i,j) \in F^+} \ln \frac{W_{ij}}{W_{ji}} = \Delta \mu - k_B T f.
    \label{DBcycle}
\end{equation}

\begin{figure}[t]
\includegraphics[width=\columnwidth]{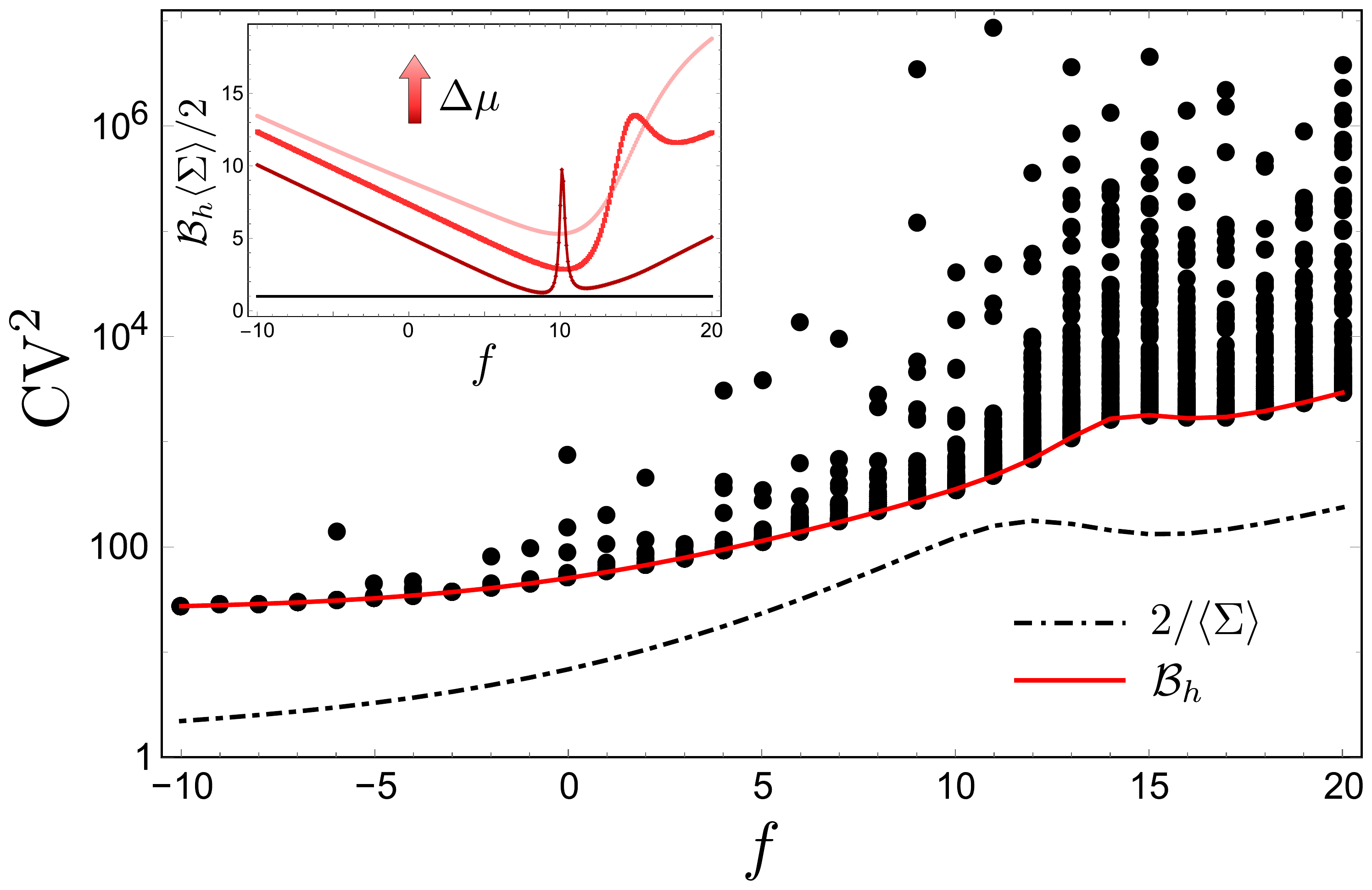}
\caption{Hyperaccurate bound for kinesin. \textit{Main}: Hyperaccurate bound (solid red) and TUR (dot-dashed black) as a function of the load force $f$ for $\Delta \mu=14.73$. The values of the all parameters have been fixed as in Ref. \cite{liepelt1}. Black dots are random currents ($10^2$ for each $f$). \textit{Inset}: The ratio between the two bounds, $B_h \langle \Sigma \rangle / 2$ is shown for $\Delta \mu = 19.34$ (pink), $14.73$ (red), and $10.12$ (dark red), corresponding to $[{\rm ATP}] = 10^{-10} M, 10^{-8} M$ and $10^{-6} M$, respectively, with the concentrations of ADP and P fixed to $50 \mu M$. The black line indicates $1$. The moving peak corresponds to $f = \Delta \mu$, with $k_B T = 1$, so that the kinesin can only dissipate energy, without performing net motion.}
\label{fig2}
\end{figure}

We notice that $K_{\rm eq}$ in Eq.~\eqref{Dmu} can be written as:
\begin{equation}
    K_{\rm eq} = \frac{\kappa_{52}\kappa_{21}\kappa_{65}\kappa_{16}}{\kappa_{25}\kappa_{12}\kappa_{56}\kappa_{61}}=\frac{\kappa_{25}\kappa_{54}\kappa_{32}\kappa_{43}}{\kappa_{52}\kappa_{45}\kappa_{23}\kappa_{34}}.
\end{equation}
From the transition matrix, hyperaccurate coefficients and bound can be readily
obtained by employing the framework outlined above. In Fig.~\ref{fig2}, we report $\mathcal{B}_h$ (solid red) together with the bound provided by the TUR (dot-dashed black) and an ensemble of $\CVSQ$ of random currents (black dots), as a function of the dimensionless load force $f$. By construction, $\mathcal{B}_h$ provides the tighest possible bound to the $\CVSQ$ and is markedly tighter than $2/\langle \Sigma \rangle$. Indeed, we also report in the inset the ratio $\mathcal{B}_h \langle \Sigma \rangle / 2$ for different values of $\Delta \mu$, which quantifies the difference between $B_h$ and the TUR bound. As the system approaches equilibrium, this ratio decreases, as expected \cite{pigolotti2017generic,busiello2019hyperaccurate}. Moreover, its behavior as a function of $f$ is non-monotonous, exhibiting also the presence of a peak corresponding to the value $k_B T f = \Delta \mu$, where mechanical cycles become futile and the motor only dissipates energy (see Eq.~\eqref{DBcycle}).

\begin{figure*}[th]
%\begin{figure*}[t]
%\includegraphics[width=\columnwidth]{Fig3.pdf}
\includegraphics[scale=0.3]{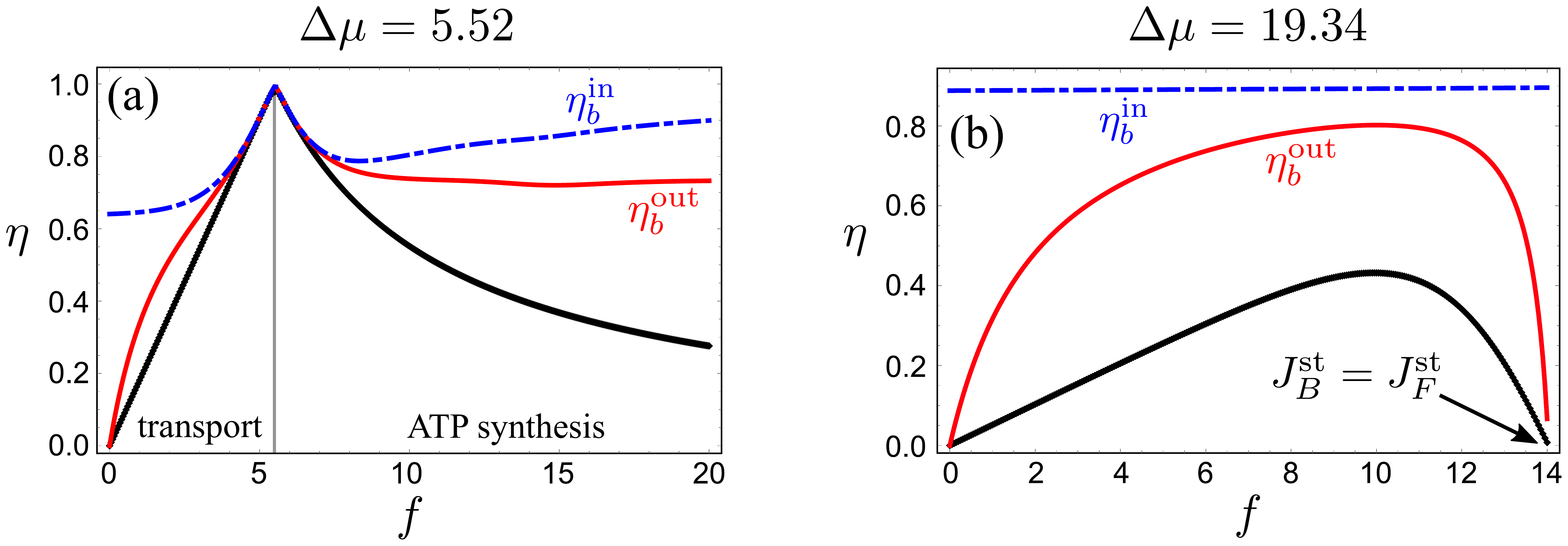}
\caption{Hyperaccurate efficiency bound for kinesin. $(a)$ The
efficiency (thick black) and the hyperaccurate bounds given by Eqs.~\eqref{bound} (solid red) and \eqref{bound2} (dot-dashed blue) are shown for $\Delta \mu = 5.52$ ($[{\rm ATP}] = 10^{-12}$). The peak corresponds to the value of $f$ for which the kinesin changes behavior, from a transporter (before the peak) to an ATP synthesizer (after the peak). $(b)$ The same comparison is presented for $\Delta \mu = 19.34$. The kinesin stops moving forward when $J^{\rm st}_B = J^{\rm st}_F$, and $\eta$ vanishes.}
\label{fig3}
\end{figure*}

To quantify the performance of the hyperaccurate efficiency bounds, we explicitly write down the steady-state entropy production.
From Eq.~\eqref{ep}, $\langle \Sigma \rangle$ reads
%\begin{equation}
%\CEF{\langle \Sigma \rangle = (2J^{{\rm st}}_B - J^{{\rm st}}_F) \frac{\Delta %\mu}{k_B T} - J^{{\rm st}}_F  f,}
%\label{ep}
%\end{equation}
\begin{equation}
\langle \Sigma \rangle = (J^{{\rm st}}_F + J^{{\rm st}}_B) \Delta \mu - (J^{{\rm st}}_F - J^{{\rm st}}_B) k_B T f,
\label{ep}
\end{equation}
where $J^{{\rm st}}_B$ and $J^{{\rm st}}_F$ are the steady-state probability fluxes associated with the cycles $B^+$ and $F^+$, respectively. Notice that $J^{{\rm st}}_F + J^{{\rm st}}_B = J^{{\rm st}}_{16} + J^{{\rm st}}_{43}$, which is the total thermodynamic flux associated with ATP hydrolysis, $\Delta \mu$. Analogously, $J^{{\rm st}}_F - J^{\rm st}_B = J^{\rm st}_{52}$ is associated with the mechanical step, and thus with (minus) the load force $f$.

For a given force $f$, large values of $\Delta \mu$ allow the kinesin to work as a motor, converting chemical energy into mechanical motion. However, when $\Delta \mu$ is small, the available energy is not sufficient to displace the kinesin, hence it effectively uses mechanical energy to produce ATP or ADP (depending on the sign of $f$ and $\Delta \mu$) \cite{altaner2015fluctuating}. For simplicity, we perform the numerical analysis for $f > 0$ and $\Delta \mu > 0$, although it can be straightforwardly extended to all other cases.
From Eq.~\eqref{eff}, when the kinesin converts chemical into mechanical energy, the efficiency reads
\begin{equation}
    \eta = \frac{J^{{\rm st}}_F - J^{{\rm st}}_B}{J^{{\rm st}}_F + J^{{\rm st}}_B} \frac{k_B T f}{\Delta \mu},
\end{equation}
where the numerator is the output work $W_{\rm out} = -(J^{{\rm st}}_F - J^{{\rm st}}_B) k_B T f$, since kinesin operates against the external load force. Fig.~\ref{fig3} shows the efficiency and its associated bounds for two representative values of $\Delta \mu$. Results for other values of $\Delta \mu$ (not shown) exhibit similar features.
For $\Delta \mu = 5.52$ (Fig.~\ref{fig3}a), $\eta_b^{\rm out}$ provides a very tight bound for small values of $f$, while both $\eta_b^{\rm out}$ and $\eta_b^{\rm in}$ converges to the actual efficiency when $\eta$ approaches its maximum. We also report a change of behavior of the kinesin before and after the maximum efficiency, highlighting the change of regime from molecular transporter to an ATP synthesizer,
respectively. In these conditions, high efficiencies are also associated with small fluxes since for small $\Delta \mu$ the system is close to equilibrium. Conversely, when $\Delta \mu = 19.34$ (Fig.~\ref{fig3}b), the system is far from equilibrium and exhibits large probability fluxes. Although hyperaccurate efficiency bounds are less tight than for $\Delta \mu=5.52$, $\eta_b^{\rm out}$ still
provides a tighter bound for the efficiency than the one derived using TUR \cite{pietzonka2016universal2}. We also notice that the system stops operating as a work-to-work converter when the flux in the backward cycle is equal to the one in the forward cycle, i.e., $J^{\rm st}_B = J^{\rm st}_F$. %As a last comment, we highlight that, by construction, the knowledge of $\mathcal{B}_h$ leads to a tighter bound than TUR on the efficiency \cite{pietzonka2016universal2}.

\subsection{Hyperaccurate currents and dissipation-driven selection of states}

As a second application, we study a three-state chemical system diffusing in a temperature gradient. This model has been introduced in Ref. \cite{busiello2021dissipation} as a paradigmatic example of a selection of chemical states driven by internal dissipation processes. Later on, a possible solution to the furanose conundrum has been proposed starting from an analogous modelization \cite{dass2021equilibrium}. These studies have been stimulated by, and in turn fueled, the idea that life might have been an inevitable consequence of nonequilibrium thermodynamics \cite{busiello2021dissipation}.

\begin{figure}[t]
\includegraphics[width=\columnwidth]{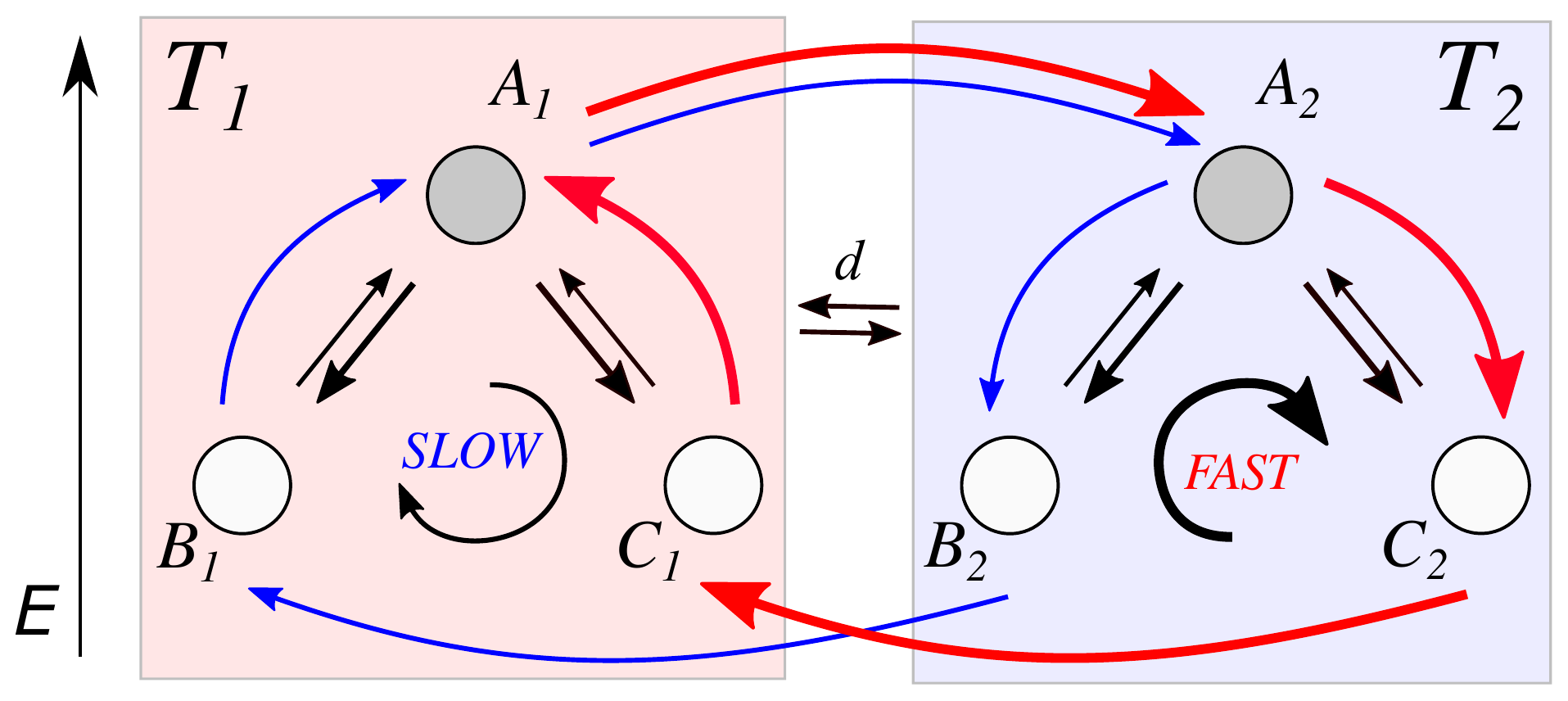}
\caption{Three-state chemical system in a temperature gradient, $T_1 > T_2$. Each state can diffuse between the boxes with the same diffusion rate, $d$, and $A$ can convert into $B$ or $C$ in both boxes. The vertical position of a state is proportional to its energy, i.e., $E_A > E_B = E_C$. Red arrows indicate the net steady-state probability flux flowing through $C$ states only. This is the fastest net flux in the system. Blue arrows are associated to the slow net flux that only visit $B$ states.}
\label{fig5}
\end{figure}

The system consists of three chemical states, $A, B$ and $C$, living in two different boxes at two different temperatures, $T_1$ and $T_2$ with $T_1 > T_2$. Moreover, each chemical species can diffusively move between the boxes, leading to a 6-state model, as depicted in Fig.~\ref{fig5}. All possible internal transitions among states are:
\begin{equation}
    A_i \leftrightharpoons B_i \qquad A_i \leftrightharpoons C_i \qquad i = 1, 2
\end{equation}
where $X_i$ indicates the species $X$ in the box $i$. 

To determine the transition matrix governing the system dynamics, we write the chemical rates in the standard Kramers' form \cite{busiello2021dissipation}:
\begin{eqnarray}
k_{B_i A_i} = e^{-\Delta E/k_B T_i} k_{A_i B_i} \nonumber \\
k_{C_i A_i} = e^{-\Delta E/k_B T_i} k_{A_i C_i}
\end{eqnarray}
where $k_{X_i Y_i}$ is the chemical rates associated with the reaction from $X_i$ to $Y_i$, and $\Delta E = E_A - E_B = E_A - E_C$ for simplicity. We introduce a kinetic asymmetry by setting two different energetic barriers in going from $A_i$ to $B_i$, $\Delta \epsilon_B$, and from $A_i$ to $C_i$, $\Delta \epsilon_C$, so that:
\begin{equation}
k_{A_i B_i} = e^{- \Delta \epsilon /k_B T_i} k_{A_i C_i},
\end{equation}
with $\Delta \epsilon = \Delta \epsilon_B - \Delta \epsilon_C > 0$, which means that the state $C$ is kinetically favorable with respect to $B$ \cite{busiello2021dissipation}.

We also assume that all species move between boxes with the same (symmetric) diffusive rate, $d_A = d_B = d_C = d$, for simplicity. We are interested in determining the total population of species $B$ and $C$ at stationarity, i.e., $[B_1]^{\rm st} + [B_2]^{\rm st} := [B]^{\rm st}$ and $[C_1]^{\rm st} + [C_2]^{\rm st} := [C]^{\rm st}$, respectively. To quantify the unbalance between these two, we introduce the selection parameter $R_{CB} = \log ([C]^{\rm st}/[B]^{\rm st})$, which can be interpreted as the stabilization energy of $C$ with respect to $B$.
When $T_1 = T_2 = T$, the system is at thermodynamic equilibrium and eventually reaches a Boltzmann distribution in which  the states $B$ and $C$ are equally populated since they have the same energy. Analogously, when $d = 0$, each box will relax to its own Boltzmann distribution with temperature $T_i$, and the total population of $B$ will be identical to the one of $C$. However, in the presence of diffusion and a temperature gradient, the system dissipates thermal energy performing diffusive cycles between boxes. In particular, two stationary fluxes emerge: one only flows through $B$ states (blue arrows in Fig.~\ref{fig5}) and exhibits slow dissipation, while the other only flows through $C$ states and dissipates faster (red arrows in Fig.~\ref{fig5}). This symmetry breaking is associated with the kinetic asymmetry in the energetic barriers and will result in a steady-state population $[C]^{\rm st}$ higher than $[B]^{\rm st}$, i.e., $R_{CB} > 1$.

\begin{figure}[t]
\includegraphics[width=\columnwidth]{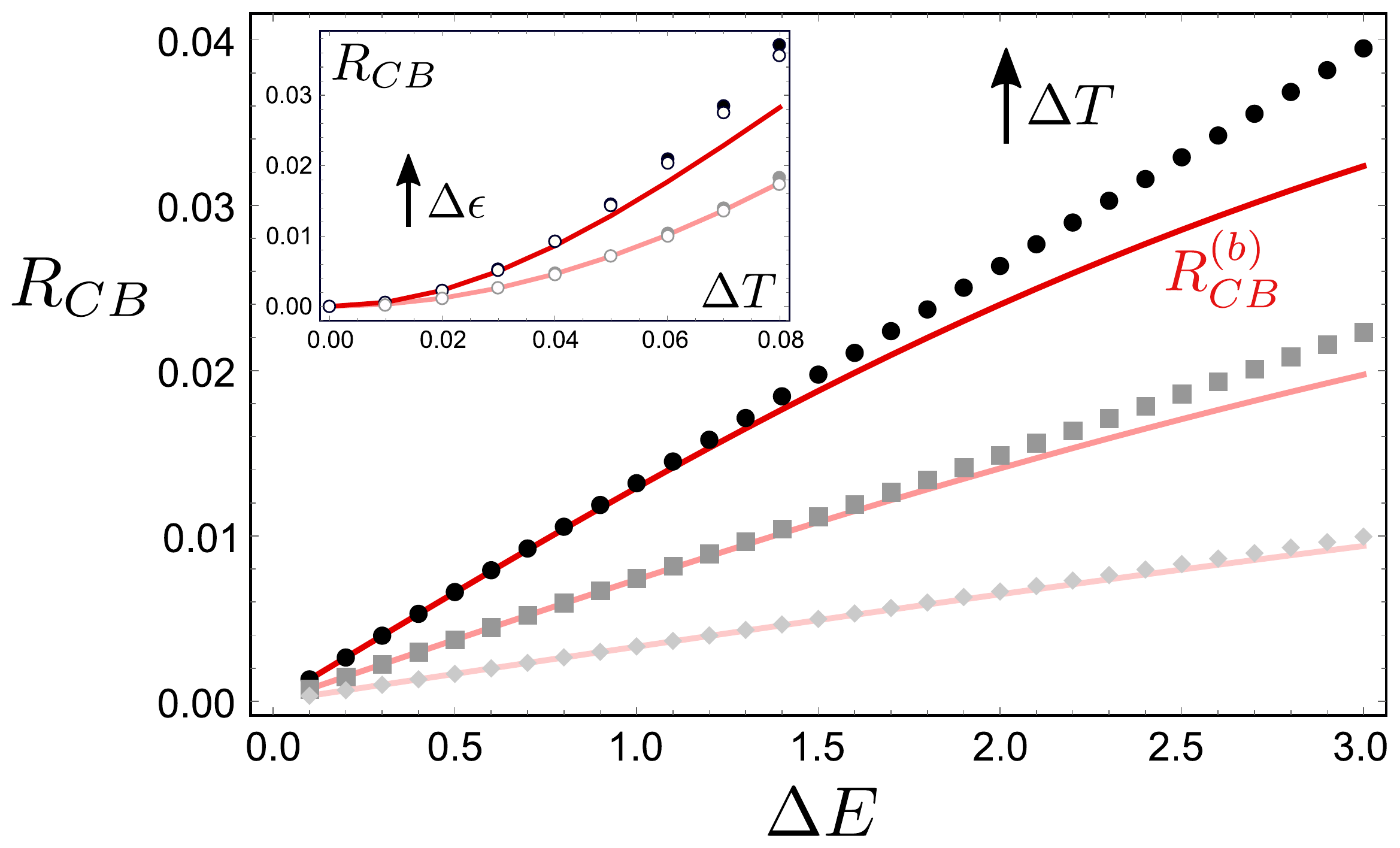}
\caption{Hyperaccurate bound for dissipation-driven selection of states. \textit{Main}: For increasing value of $\Delta T \in [0.04,0.08]$ (with decreasing opacity), the selection parameter $R_{CB}$ is shown as a function of $\Delta E$ (points), along with the bound $R_{CB}^{(h)}$ (red curves), for $\Delta \epsilon_B = 3$, $\Delta \epsilon_C = 1$, $T_m = 0.7$. \textit{Inset}: $R_{CB}$ (open markers), its approximation for small $\Delta T$ in Eq.~\eqref{RS} (filled markers), and its bound  $R_{CB}^{(b)}$(red curves) are reported as a function of $\Delta T$ for increasing value of the kinetic asymmetry, $\Delta \epsilon$ (with decreasing opacity). Here, $\Delta \epsilon_B = 3$, $\Delta \epsilon_C \in [0.2,1.6]$, and $T_m = 0.7$. In all plots, we set $d \to +\infty$.}
\label{fig6}
\end{figure}

It is possible to show that, in the limit of fast diffusion $d \to +\infty$, and small gradient $T_1 \gtrsim T_2$, we have:
\begin{equation}
    R_{CB} = \log\left(1 + \frac{\langle \Sigma \rangle}{\Delta E} \frac{1}{P^{\rm eq}_m(A)} \frac{\Delta \epsilon}{e^{-\frac{\Delta\epsilon_B}{k_B T_m}} + e^{-\frac{\Delta\epsilon_C}{k_B T_m}}}\right)
    \label{RS}
\end{equation}
where $P^{\rm eq}_m(A)$ is the equilibrium probability distribution of $A$ at the average temperature $T_m = (T_1 + T_2)/2$ and $\langle\Sigma\rangle$ is the entropy production. Moreover, a positive correlation between $R_{CB}$ and $\langle\Sigma \rangle$ has been shown to hold even beyond this limit \cite{busiello2021dissipation}.

Providing a lower bound for the average entropy production, $\langle \Sigma \rangle$, we obtain a lower bound for $R_{CB}$, namely $R_{CB}^{(b)}$, in terms of the hyperaccurate bound.
%, by means of a bound for $\langle \Sigma \rangle$. 
In Fig.~\ref{fig5}, we show $R_{CB}$ and $R_{CB}^{(b)}$ as a function of $\Delta E$, for increasing values of the temperature gradient. When $\Delta E$ is small, our bound predicts the correct value of selection, whereas some deviations appears when $\Delta E$ increases. Similar findings are reported in the inset, in which the selection parameter (open circles), its estimation from Eq.~\eqref{RS} (black dots), and the bound here derived (in red) are reported versus $\Delta T$ for two different values of the kinetic asymmetry, $\Delta \epsilon$. Once again, the bound obtained from the hyperaccurate current, $R_{CB}^{(b)}$, provides an accurate prediction for the selection parameter close to the equilibrium, while deviations arise as the temperature gradient increases.

\section{Conclusions}

Thermodynamic uncertainty relations (TURs) set universal bounds for the \textit{precision} of a stochastic current, quantified as the ratio between its variance and squared mean ($\CVSQ$), in terms of the dissipation of the system. Thus, by inverting this inequality, it is possible to provide a lower bound to the entropy production and estimate the distance from thermodynamic equilibrium. The main advantage of this indirect approach is that it does not require the large sample sizes and observational times that are commonly required to provide a direct estimation of the entropy production. In \cite{busiello2019hyperaccurate}, it has been pointed out how the knowledge of the hyperaccurate current, i.e., the one with the minimum $\CVSQ$, might greatly improve our predicting power. Here, we introduced and derived the hyperaccurate current for generic discrete-state Markovian systems, both evolving in discrete and continuous time. Our analytical closed formula has been tested against two models of chemical systems. As a side result, we also provided two hyperaccurate bounds for the efficiency of work-to-work converters, as a function of either the input or the output power. Possible future extensions might include the study of nonintegrated currents and nonstationary dynamics, and a connection among TURs, hyperaccuracy, and information theory, whose role is becoming dominant in understanding stochastic systems \cite{ito2018stochastic,nicoletti2021mutual}.

Additionally, we employed our framework to compute the finite-time hyperaccurate bounds for one-loop discrete-state systems (rings).
%that one-loop systems do not have an hyperaccurate current in the long-time limit. However, since our framework can be easily extended to finite times, we computed the finite-time hyperaccurate bounds for these cases. 
Our results suggest that short-time experiments might be much more informative than the long-time limit to estimate the average dissipation. A formal proof and generalization of this statement, along with a feasible experimental approach for discrete-state Markovian dynamics, is left as an intriguing perspective for the future.

\section{Acknowledgments}
We gratefully acknowledge Simone Pigolotti for insightful suggestions, inspiring discussions, and the help in developing the idea. C.E.F. acknowledges the financial support from Santander call "New international
partners" under Grant 1145/2019 and São
Paulo Research Foundation (FAPESP) under Grants
 No. 2021/05503-7 and  2021/03372-2. The financial support from CNPq is also acknowledged.

%\bibliography{refs}

\begin{thebibliography}{52}
\makeatletter
\providecommand \@ifxundefined [1]{%
 \@ifx{#1\undefined}
}%
\providecommand \@ifnum [1]{%
 \ifnum #1\expandafter \@firstoftwo
 \else \expandafter \@secondoftwo
 \fi
}%
\providecommand \@ifx [1]{%
 \ifx #1\expandafter \@firstoftwo
 \else \expandafter \@secondoftwo
 \fi
}%
\providecommand \natexlab [1]{#1}%
\providecommand \enquote  [1]{``#1''}%
\providecommand \bibnamefont  [1]{#1}%
\providecommand \bibfnamefont [1]{#1}%
\providecommand \citenamefont [1]{#1}%
\providecommand \href@noop [0]{\@secondoftwo}%
\providecommand \href [0]{\begingroup \@sanitize@url \@href}%
\providecommand \@href[1]{\@@startlink{#1}\@@href}%
\providecommand \@@href[1]{\endgroup#1\@@endlink}%
\providecommand \@sanitize@url [0]{\catcode `\\12\catcode `\$12\catcode
  `\&12\catcode `\#12\catcode `\^12\catcode `\_12\catcode `\%12\relax}%
\providecommand \@@startlink[1]{}%
\providecommand \@@endlink[0]{}%
\providecommand \url  [0]{\begingroup\@sanitize@url \@url }%
\providecommand \@url [1]{\endgroup\@href {#1}{\urlprefix }}%
\providecommand \urlprefix  [0]{URL }%
\providecommand \Eprint [0]{\href }%
\providecommand \doibase [0]{http://dx.doi.org/}%
\providecommand \selectlanguage [0]{\@gobble}%
\providecommand \bibinfo  [0]{\@secondoftwo}%
\providecommand \bibfield  [0]{\@secondoftwo}%
\providecommand \translation [1]{[#1]}%
\providecommand \BibitemOpen [0]{}%
\providecommand \bibitemStop [0]{}%
\providecommand \bibitemNoStop [0]{.\EOS\space}%
\providecommand \EOS [0]{\spacefactor3000\relax}%
\providecommand \BibitemShut  [1]{\csname bibitem#1\endcsname}%
\let\auto@bib@innerbib\@empty
%</preamble>
\bibitem [{\citenamefont {De~Groot}\ and\ \citenamefont
  {Mazur}(2013)}]{de2013non}%
  \BibitemOpen
  \bibfield  {author} {\bibinfo {author} {\bibfnamefont {S.~R.}\ \bibnamefont
  {De~Groot}}\ and\ \bibinfo {author} {\bibfnamefont {P.}~\bibnamefont
  {Mazur}},\ }\href@noop {} {\emph {\bibinfo {title} {Non-equilibrium
  thermodynamics}}}\ (\bibinfo  {publisher} {Courier Corporation},\ \bibinfo
  {year} {2013})\BibitemShut {NoStop}%
\bibitem [{\citenamefont {Seifert}(2012)}]{seifert2012stochastic}%
  \BibitemOpen
  \bibfield  {author} {\bibinfo {author} {\bibfnamefont {U.}~\bibnamefont
  {Seifert}},\ }\href@noop {} {\bibfield  {journal} {\bibinfo  {journal}
  {Reports on progress in physics}\ }\textbf {\bibinfo {volume} {75}},\
  \bibinfo {pages} {126001} (\bibinfo {year} {2012})}\BibitemShut {NoStop}%
\bibitem [{\citenamefont {Ciliberto}(2017)}]{ciliberto2017experiments}%
  \BibitemOpen
  \bibfield  {author} {\bibinfo {author} {\bibfnamefont {S.}~\bibnamefont
  {Ciliberto}},\ }\href@noop {} {\bibfield  {journal} {\bibinfo  {journal}
  {Physical Review X}\ }\textbf {\bibinfo {volume} {7}},\ \bibinfo {pages}
  {021051} (\bibinfo {year} {2017})}\BibitemShut {NoStop}%
\bibitem [{\citenamefont {Tom{\'e}}\ and\ \citenamefont
  {de~Oliveira}(2015)}]{tome2015stochastic}%
  \BibitemOpen
  \bibfield  {author} {\bibinfo {author} {\bibfnamefont {T.}~\bibnamefont
  {Tom{\'e}}}\ and\ \bibinfo {author} {\bibfnamefont {M.~J.}\ \bibnamefont
  {de~Oliveira}},\ }\href@noop {} {\bibfield  {journal} {\bibinfo  {journal}
  {Physical review E}\ }\textbf {\bibinfo {volume} {91}},\ \bibinfo {pages}
  {042140} (\bibinfo {year} {2015})}\BibitemShut {NoStop}%
\bibitem [{\citenamefont {Van~den Broeck}\ and\ \citenamefont
  {Esposito}(2015)}]{van2015ensemble}%
  \BibitemOpen
  \bibfield  {author} {\bibinfo {author} {\bibfnamefont {C.}~\bibnamefont
  {Van~den Broeck}}\ and\ \bibinfo {author} {\bibfnamefont {M.}~\bibnamefont
  {Esposito}},\ }\href@noop {} {\bibfield  {journal} {\bibinfo  {journal}
  {Physica A: Statistical Mechanics and its Applications}\ }\textbf {\bibinfo
  {volume} {418}},\ \bibinfo {pages} {6} (\bibinfo {year} {2015})}\BibitemShut
  {NoStop}%
\bibitem [{\citenamefont {Akasaki}\ \emph {et~al.}(2020)\citenamefont
  {Akasaki}, \citenamefont {de~Oliveira},\ and\ \citenamefont
  {Fiore}}]{akasaki}%
  \BibitemOpen
  \bibfield  {author} {\bibinfo {author} {\bibfnamefont {B.~A.~N.}\
  \bibnamefont {Akasaki}}, \bibinfo {author} {\bibfnamefont {M.~J.}\
  \bibnamefont {de~Oliveira}}, \ and\ \bibinfo {author} {\bibfnamefont {C.~E.}\
  \bibnamefont {Fiore}},\ }\href@noop {} {\bibfield  {journal} {\bibinfo
  {journal} {Phys. Rev. E}\ }\textbf {\bibinfo {volume} {101}},\ \bibinfo
  {pages} {012132} (\bibinfo {year} {2020})}\BibitemShut {NoStop}%
\bibitem [{\citenamefont {Tom\'e}\ and\ \citenamefont
  {de~Oliveira}(2010)}]{tome2010}%
  \BibitemOpen
  \bibfield  {author} {\bibinfo {author} {\bibfnamefont {T.}~\bibnamefont
  {Tom\'e}}\ and\ \bibinfo {author} {\bibfnamefont {M.~J.}\ \bibnamefont
  {de~Oliveira}},\ }\href@noop {} {\bibfield  {journal} {\bibinfo  {journal}
  {Physical review E}\ }\textbf {\bibinfo {volume} {82}},\ \bibinfo {pages}
  {021120} (\bibinfo {year} {2010})}\BibitemShut {NoStop}%
\bibitem [{\citenamefont {Rao}\ and\ \citenamefont
  {Esposito}(2016)}]{rao2016nonequilibrium}%
  \BibitemOpen
  \bibfield  {author} {\bibinfo {author} {\bibfnamefont {R.}~\bibnamefont
  {Rao}}\ and\ \bibinfo {author} {\bibfnamefont {M.}~\bibnamefont {Esposito}},\
  }\href@noop {} {\bibfield  {journal} {\bibinfo  {journal} {Physical Review
  X}\ }\textbf {\bibinfo {volume} {6}},\ \bibinfo {pages} {041064} (\bibinfo
  {year} {2016})}\BibitemShut {NoStop}%
\bibitem [{\citenamefont {Esposito}(2020)}]{esposito2020open}%
  \BibitemOpen
  \bibfield  {author} {\bibinfo {author} {\bibfnamefont {M.}~\bibnamefont
  {Esposito}},\ }\href@noop {} {\bibfield  {journal} {\bibinfo  {journal}
  {Communications Chemistry}\ }\textbf {\bibinfo {volume} {3}},\ \bibinfo
  {pages} {1} (\bibinfo {year} {2020})}\BibitemShut {NoStop}%
\bibitem [{\citenamefont {Busiello}\ \emph {et~al.}(2021)\citenamefont
  {Busiello}, \citenamefont {Liang}, \citenamefont {Piazza},\ and\
  \citenamefont {De~Los~Rios}}]{busiello2021dissipation}%
  \BibitemOpen
  \bibfield  {author} {\bibinfo {author} {\bibfnamefont {D.~M.}\ \bibnamefont
  {Busiello}}, \bibinfo {author} {\bibfnamefont {S.}~\bibnamefont {Liang}},
  \bibinfo {author} {\bibfnamefont {F.}~\bibnamefont {Piazza}}, \ and\ \bibinfo
  {author} {\bibfnamefont {P.}~\bibnamefont {De~Los~Rios}},\ }\href@noop {}
  {\bibfield  {journal} {\bibinfo  {journal} {Communications Chemistry}\
  }\textbf {\bibinfo {volume} {4}},\ \bibinfo {pages} {1} (\bibinfo {year}
  {2021})}\BibitemShut {NoStop}%
\bibitem [{\citenamefont {Noa}\ \emph {et~al.}(2019)\citenamefont {Noa},
  \citenamefont {Harunari}, \citenamefont {de~Oliveira},\ and\ \citenamefont
  {Fiore}}]{PhysRevE.100.012104}%
  \BibitemOpen
  \bibfield  {author} {\bibinfo {author} {\bibfnamefont {C.~E.~F.}\
  \bibnamefont {Noa}}, \bibinfo {author} {\bibfnamefont {P.~E.}\ \bibnamefont
  {Harunari}}, \bibinfo {author} {\bibfnamefont {M.~J.}\ \bibnamefont
  {de~Oliveira}}, \ and\ \bibinfo {author} {\bibfnamefont {C.~E.}\ \bibnamefont
  {Fiore}},\ }\href@noop {} {\bibfield  {journal} {\bibinfo  {journal} {Phys.
  Rev. E}\ }\textbf {\bibinfo {volume} {100}},\ \bibinfo {pages} {012104}
  (\bibinfo {year} {2019})}\BibitemShut {NoStop}%
\bibitem [{\citenamefont {Nguyen}\ and\ \citenamefont
  {Seifert}(2020)}]{PhysRevE.102.022101}%
  \BibitemOpen
  \bibfield  {author} {\bibinfo {author} {\bibfnamefont {B.}~\bibnamefont
  {Nguyen}}\ and\ \bibinfo {author} {\bibfnamefont {U.}~\bibnamefont
  {Seifert}},\ }\href@noop {} {\bibfield  {journal} {\bibinfo  {journal} {Phys.
  Rev. E}\ }\textbf {\bibinfo {volume} {102}},\ \bibinfo {pages} {022101}
  (\bibinfo {year} {2020})}\BibitemShut {NoStop}%
\bibitem [{\citenamefont {Fiore}\ \emph {et~al.}(2021)\citenamefont {Fiore},
  \citenamefont {Harunari}, \citenamefont {Noa},\ and\ \citenamefont
  {Landi}}]{PhysRevE.104.064123}%
  \BibitemOpen
  \bibfield  {author} {\bibinfo {author} {\bibfnamefont {C.~E.}\ \bibnamefont
  {Fiore}}, \bibinfo {author} {\bibfnamefont {P.~E.}\ \bibnamefont {Harunari}},
  \bibinfo {author} {\bibfnamefont {C.~E.~F.}\ \bibnamefont {Noa}}, \ and\
  \bibinfo {author} {\bibfnamefont {G.~T.}\ \bibnamefont {Landi}},\ }\href@noop
  {} {\bibfield  {journal} {\bibinfo  {journal} {Phys. Rev. E}\ }\textbf
  {\bibinfo {volume} {104}},\ \bibinfo {pages} {064123} (\bibinfo {year}
  {2021})}\BibitemShut {NoStop}%
\bibitem [{\citenamefont {Proesmans}\ \emph {et~al.}(2016)\citenamefont
  {Proesmans}, \citenamefont {Cleuren},\ and\ \citenamefont {Van~den
  Broeck}}]{proesmans2016linear}%
  \BibitemOpen
  \bibfield  {author} {\bibinfo {author} {\bibfnamefont {K.}~\bibnamefont
  {Proesmans}}, \bibinfo {author} {\bibfnamefont {B.}~\bibnamefont {Cleuren}},
  \ and\ \bibinfo {author} {\bibfnamefont {C.}~\bibnamefont {Van~den Broeck}},\
  }\href@noop {} {\bibfield  {journal} {\bibinfo  {journal} {Journal of
  Statistical Mechanics: Theory and Experiment}\ }\textbf {\bibinfo {volume}
  {2016}},\ \bibinfo {pages} {023202} (\bibinfo {year} {2016})}\BibitemShut
  {NoStop}%
\bibitem [{\citenamefont {Brandner}\ \emph {et~al.}(2015)\citenamefont
  {Brandner}, \citenamefont {Saito},\ and\ \citenamefont
  {Seifert}}]{brandner2015thermodynamics}%
  \BibitemOpen
  \bibfield  {author} {\bibinfo {author} {\bibfnamefont {K.}~\bibnamefont
  {Brandner}}, \bibinfo {author} {\bibfnamefont {K.}~\bibnamefont {Saito}}, \
  and\ \bibinfo {author} {\bibfnamefont {U.}~\bibnamefont {Seifert}},\
  }\href@noop {} {\bibfield  {journal} {\bibinfo  {journal} {Physical review
  X}\ }\textbf {\bibinfo {volume} {5}},\ \bibinfo {pages} {031019} (\bibinfo
  {year} {2015})}\BibitemShut {NoStop}%
\bibitem [{\citenamefont {Proesmans}\ and\ \citenamefont {Van~den
  Broeck}(2015)}]{proesmans2015onsager}%
  \BibitemOpen
  \bibfield  {author} {\bibinfo {author} {\bibfnamefont {K.}~\bibnamefont
  {Proesmans}}\ and\ \bibinfo {author} {\bibfnamefont {C.}~\bibnamefont
  {Van~den Broeck}},\ }\href@noop {} {\bibfield  {journal} {\bibinfo  {journal}
  {Physical review letters}\ }\textbf {\bibinfo {volume} {115}},\ \bibinfo
  {pages} {090601} (\bibinfo {year} {2015})}\BibitemShut {NoStop}%
\bibitem [{\citenamefont {Proesmans}\ and\ \citenamefont
  {Fiore}(2019)}]{fiorek}%
  \BibitemOpen
  \bibfield  {author} {\bibinfo {author} {\bibfnamefont {K.}~\bibnamefont
  {Proesmans}}\ and\ \bibinfo {author} {\bibfnamefont {C.~E.}\ \bibnamefont
  {Fiore}},\ }\href@noop {} {\bibfield  {journal} {\bibinfo  {journal} {Phys.
  Rev. E}\ }\textbf {\bibinfo {volume} {100}},\ \bibinfo {pages} {022141}
  (\bibinfo {year} {2019})}\BibitemShut {NoStop}%
\bibitem [{\citenamefont {Busiello}\ \emph {et~al.}(2020)\citenamefont
  {Busiello}, \citenamefont {Gupta},\ and\ \citenamefont
  {Maritan}}]{busiello2020coarse}%
  \BibitemOpen
  \bibfield  {author} {\bibinfo {author} {\bibfnamefont {D.~M.}\ \bibnamefont
  {Busiello}}, \bibinfo {author} {\bibfnamefont {D.}~\bibnamefont {Gupta}}, \
  and\ \bibinfo {author} {\bibfnamefont {A.}~\bibnamefont {Maritan}},\
  }\href@noop {} {\bibfield  {journal} {\bibinfo  {journal} {Physical Review
  Research}\ }\textbf {\bibinfo {volume} {2}},\ \bibinfo {pages} {043257}
  (\bibinfo {year} {2020})}\BibitemShut {NoStop}%
\bibitem [{\citenamefont {Raz}\ \emph {et~al.}(2016)\citenamefont {Raz},
  \citenamefont {Suba{\c{s}}{\i}},\ and\ \citenamefont
  {Jarzynski}}]{raz2016mimicking}%
  \BibitemOpen
  \bibfield  {author} {\bibinfo {author} {\bibfnamefont {O.}~\bibnamefont
  {Raz}}, \bibinfo {author} {\bibfnamefont {Y.}~\bibnamefont
  {Suba{\c{s}}{\i}}}, \ and\ \bibinfo {author} {\bibfnamefont {C.}~\bibnamefont
  {Jarzynski}},\ }\href@noop {} {\bibfield  {journal} {\bibinfo  {journal}
  {Physical Review X}\ }\textbf {\bibinfo {volume} {6}},\ \bibinfo {pages}
  {021022} (\bibinfo {year} {2016})}\BibitemShut {NoStop}%
\bibitem [{\citenamefont {Busiello}\ \emph {et~al.}(2018)\citenamefont
  {Busiello}, \citenamefont {Jarzynski},\ and\ \citenamefont
  {Raz}}]{busiello2018similarities}%
  \BibitemOpen
  \bibfield  {author} {\bibinfo {author} {\bibfnamefont {D.~M.}\ \bibnamefont
  {Busiello}}, \bibinfo {author} {\bibfnamefont {C.}~\bibnamefont {Jarzynski}},
  \ and\ \bibinfo {author} {\bibfnamefont {O.}~\bibnamefont {Raz}},\
  }\href@noop {} {\bibfield  {journal} {\bibinfo  {journal} {New Journal of
  Physics}\ }\textbf {\bibinfo {volume} {20}},\ \bibinfo {pages} {093015}
  (\bibinfo {year} {2018})}\BibitemShut {NoStop}%
\bibitem [{\citenamefont {Jarzynski}(1997)}]{jarzynski1997nonequilibrium}%
  \BibitemOpen
  \bibfield  {author} {\bibinfo {author} {\bibfnamefont {C.}~\bibnamefont
  {Jarzynski}},\ }\href@noop {} {\bibfield  {journal} {\bibinfo  {journal}
  {Physical Review Letters}\ }\textbf {\bibinfo {volume} {78}},\ \bibinfo
  {pages} {2690} (\bibinfo {year} {1997})}\BibitemShut {NoStop}%
\bibitem [{\citenamefont {Li}\ \emph {et~al.}(2019)\citenamefont {Li},
  \citenamefont {Horowitz}, \citenamefont {Gingrich},\ and\ \citenamefont
  {Fakhri}}]{li2019quantifying}%
  \BibitemOpen
  \bibfield  {author} {\bibinfo {author} {\bibfnamefont {J.}~\bibnamefont
  {Li}}, \bibinfo {author} {\bibfnamefont {J.~M.}\ \bibnamefont {Horowitz}},
  \bibinfo {author} {\bibfnamefont {T.~R.}\ \bibnamefont {Gingrich}}, \ and\
  \bibinfo {author} {\bibfnamefont {N.}~\bibnamefont {Fakhri}},\ }\href@noop {}
  {\bibfield  {journal} {\bibinfo  {journal} {Nature communications}\ }\textbf
  {\bibinfo {volume} {10}},\ \bibinfo {pages} {1} (\bibinfo {year}
  {2019})}\BibitemShut {NoStop}%
\bibitem [{\citenamefont {Van~Vu}\ \emph {et~al.}(2020)\citenamefont {Van~Vu},
  \citenamefont {Hasegawa} \emph {et~al.}}]{van2020entropy}%
  \BibitemOpen
  \bibfield  {author} {\bibinfo {author} {\bibfnamefont {T.}~\bibnamefont
  {Van~Vu}}, \bibinfo {author} {\bibfnamefont {Y.}~\bibnamefont {Hasegawa}},
  \emph {et~al.},\ }\href@noop {} {\bibfield  {journal} {\bibinfo  {journal}
  {Physical Review E}\ }\textbf {\bibinfo {volume} {101}},\ \bibinfo {pages}
  {042138} (\bibinfo {year} {2020})}\BibitemShut {NoStop}%
\bibitem [{\citenamefont {Manikandan}\ \emph {et~al.}(2020)\citenamefont
  {Manikandan}, \citenamefont {Gupta},\ and\ \citenamefont
  {Krishnamurthy}}]{manikandan2020inferring}%
  \BibitemOpen
  \bibfield  {author} {\bibinfo {author} {\bibfnamefont {S.~K.}\ \bibnamefont
  {Manikandan}}, \bibinfo {author} {\bibfnamefont {D.}~\bibnamefont {Gupta}}, \
  and\ \bibinfo {author} {\bibfnamefont {S.}~\bibnamefont {Krishnamurthy}},\
  }\href@noop {} {\bibfield  {journal} {\bibinfo  {journal} {Physical review
  letters}\ }\textbf {\bibinfo {volume} {124}},\ \bibinfo {pages} {120603}
  (\bibinfo {year} {2020})}\BibitemShut {NoStop}%
\bibitem [{\citenamefont {Barato}\ and\ \citenamefont
  {Seifert}(2015)}]{barato2015thermodynamic}%
  \BibitemOpen
  \bibfield  {author} {\bibinfo {author} {\bibfnamefont {A.~C.}\ \bibnamefont
  {Barato}}\ and\ \bibinfo {author} {\bibfnamefont {U.}~\bibnamefont
  {Seifert}},\ }\href@noop {} {\bibfield  {journal} {\bibinfo  {journal}
  {Physical review letters}\ }\textbf {\bibinfo {volume} {114}},\ \bibinfo
  {pages} {158101} (\bibinfo {year} {2015})}\BibitemShut {NoStop}%
\bibitem [{\citenamefont {Horowitz}\ and\ \citenamefont
  {Gingrich}(2017)}]{PhysRevE.96.020103}%
  \BibitemOpen
  \bibfield  {author} {\bibinfo {author} {\bibfnamefont {J.~M.}\ \bibnamefont
  {Horowitz}}\ and\ \bibinfo {author} {\bibfnamefont {T.~R.}\ \bibnamefont
  {Gingrich}},\ }\href@noop {} {\bibfield  {journal} {\bibinfo  {journal}
  {Phys. Rev. E}\ }\textbf {\bibinfo {volume} {96}},\ \bibinfo {pages} {020103}
  (\bibinfo {year} {2017})}\BibitemShut {NoStop}%
\bibitem [{\citenamefont {Fischer}\ \emph {et~al.}(2020)\citenamefont
  {Fischer}, \citenamefont {Chun},\ and\ \citenamefont
  {Seifert}}]{PhysRevE.102.012120}%
  \BibitemOpen
  \bibfield  {author} {\bibinfo {author} {\bibfnamefont {L.~P.}\ \bibnamefont
  {Fischer}}, \bibinfo {author} {\bibfnamefont {H.-M.}\ \bibnamefont {Chun}}, \
  and\ \bibinfo {author} {\bibfnamefont {U.}~\bibnamefont {Seifert}},\
  }\href@noop {} {\bibfield  {journal} {\bibinfo  {journal} {Phys. Rev. E}\
  }\textbf {\bibinfo {volume} {102}},\ \bibinfo {pages} {012120} (\bibinfo
  {year} {2020})}\BibitemShut {NoStop}%
\bibitem [{\citenamefont {Hasegawa}\ and\ \citenamefont
  {Van~Vu}(2019{\natexlab{a}})}]{PhysRevE.99.062126}%
  \BibitemOpen
  \bibfield  {author} {\bibinfo {author} {\bibfnamefont {Y.}~\bibnamefont
  {Hasegawa}}\ and\ \bibinfo {author} {\bibfnamefont {T.}~\bibnamefont
  {Van~Vu}},\ }\href@noop {} {\bibfield  {journal} {\bibinfo  {journal} {Phys.
  Rev. E}\ }\textbf {\bibinfo {volume} {99}},\ \bibinfo {pages} {062126}
  (\bibinfo {year} {2019}{\natexlab{a}})}\BibitemShut {NoStop}%
\bibitem [{\citenamefont {Van~Vu}\ and\ \citenamefont
  {Hasegawa}(2019)}]{PhysRevE.100.012134}%
  \BibitemOpen
  \bibfield  {author} {\bibinfo {author} {\bibfnamefont {T.}~\bibnamefont
  {Van~Vu}}\ and\ \bibinfo {author} {\bibfnamefont {Y.}~\bibnamefont
  {Hasegawa}},\ }\href@noop {} {\bibfield  {journal} {\bibinfo  {journal}
  {Phys. Rev. E}\ }\textbf {\bibinfo {volume} {100}},\ \bibinfo {pages}
  {012134} (\bibinfo {year} {2019})}\BibitemShut {NoStop}%
\bibitem [{\citenamefont {Van~Vu}\ and\ \citenamefont
  {Hasegawa}(2020)}]{PhysRevResearch.2.013060}%
  \BibitemOpen
  \bibfield  {author} {\bibinfo {author} {\bibfnamefont {T.}~\bibnamefont
  {Van~Vu}}\ and\ \bibinfo {author} {\bibfnamefont {Y.}~\bibnamefont
  {Hasegawa}},\ }\href@noop {} {\bibfield  {journal} {\bibinfo  {journal}
  {Phys. Rev. Research}\ }\textbf {\bibinfo {volume} {2}},\ \bibinfo {pages}
  {013060} (\bibinfo {year} {2020})}\BibitemShut {NoStop}%
\bibitem [{\citenamefont {Barato}\ \emph {et~al.}(2019)\citenamefont {Barato},
  \citenamefont {Chetrite}, \citenamefont {Faggionato},\ and\ \citenamefont
  {Gabrielli}}]{barato2019unifying}%
  \BibitemOpen
  \bibfield  {author} {\bibinfo {author} {\bibfnamefont {A.}~\bibnamefont
  {Barato}}, \bibinfo {author} {\bibfnamefont {R.}~\bibnamefont {Chetrite}},
  \bibinfo {author} {\bibfnamefont {A.}~\bibnamefont {Faggionato}}, \ and\
  \bibinfo {author} {\bibfnamefont {D.}~\bibnamefont {Gabrielli}},\ }\href@noop
  {} {\bibfield  {journal} {\bibinfo  {journal} {Journal of Statistical
  Mechanics: Theory and Experiment}\ }\textbf {\bibinfo {volume} {2019}},\
  \bibinfo {pages} {084017} (\bibinfo {year} {2019})}\BibitemShut {NoStop}%
\bibitem [{\citenamefont {Proesmans}\ and\ \citenamefont {Van~den
  Broeck}(2017)}]{proesmans2017discrete}%
  \BibitemOpen
  \bibfield  {author} {\bibinfo {author} {\bibfnamefont {K.}~\bibnamefont
  {Proesmans}}\ and\ \bibinfo {author} {\bibfnamefont {C.}~\bibnamefont
  {Van~den Broeck}},\ }\href@noop {} {\bibfield  {journal} {\bibinfo  {journal}
  {EPL (Europhysics Letters)}\ }\textbf {\bibinfo {volume} {119}},\ \bibinfo
  {pages} {20001} (\bibinfo {year} {2017})}\BibitemShut {NoStop}%
\bibitem [{\citenamefont {Chiuchiu}\ and\ \citenamefont
  {Pigolotti}(2018)}]{chiuchiu2018mapping}%
  \BibitemOpen
  \bibfield  {author} {\bibinfo {author} {\bibfnamefont {D.}~\bibnamefont
  {Chiuchiu}}\ and\ \bibinfo {author} {\bibfnamefont {S.}~\bibnamefont
  {Pigolotti}},\ }\href@noop {} {\bibfield  {journal} {\bibinfo  {journal}
  {Physical Review E}\ }\textbf {\bibinfo {volume} {97}},\ \bibinfo {pages}
  {032109} (\bibinfo {year} {2018})}\BibitemShut {NoStop}%
\bibitem [{\citenamefont {Gupta}\ and\ \citenamefont
  {Busiello}(2020)}]{gupta2020tighter}%
  \BibitemOpen
  \bibfield  {author} {\bibinfo {author} {\bibfnamefont {D.}~\bibnamefont
  {Gupta}}\ and\ \bibinfo {author} {\bibfnamefont {D.~M.}\ \bibnamefont
  {Busiello}},\ }\href@noop {} {\bibfield  {journal} {\bibinfo  {journal}
  {Physical Review E}\ }\textbf {\bibinfo {volume} {102}},\ \bibinfo {pages}
  {062121} (\bibinfo {year} {2020})}\BibitemShut {NoStop}%
\bibitem [{\citenamefont {Dechant}(2018)}]{dechant2018multidimensional}%
  \BibitemOpen
  \bibfield  {author} {\bibinfo {author} {\bibfnamefont {A.}~\bibnamefont
  {Dechant}},\ }\href@noop {} {\bibfield  {journal} {\bibinfo  {journal}
  {Journal of Physics A: Mathematical and Theoretical}\ }\textbf {\bibinfo
  {volume} {52}},\ \bibinfo {pages} {035001} (\bibinfo {year}
  {2018})}\BibitemShut {NoStop}%
\bibitem [{\citenamefont {Vroylandt}\ \emph {et~al.}(2020)\citenamefont
  {Vroylandt}, \citenamefont {Proesmans},\ and\ \citenamefont
  {Gingrich}}]{vroylandt2020isometric}%
  \BibitemOpen
  \bibfield  {author} {\bibinfo {author} {\bibfnamefont {H.}~\bibnamefont
  {Vroylandt}}, \bibinfo {author} {\bibfnamefont {K.}~\bibnamefont
  {Proesmans}}, \ and\ \bibinfo {author} {\bibfnamefont {T.~R.}\ \bibnamefont
  {Gingrich}},\ }\href@noop {} {\bibfield  {journal} {\bibinfo  {journal}
  {Journal of Statistical Physics}\ }\textbf {\bibinfo {volume} {178}},\
  \bibinfo {pages} {1039} (\bibinfo {year} {2020})}\BibitemShut {NoStop}%
\bibitem [{\citenamefont {Hasegawa}\ and\ \citenamefont
  {Van~Vu}(2019{\natexlab{b}})}]{hasegawa2019fluctuation}%
  \BibitemOpen
  \bibfield  {author} {\bibinfo {author} {\bibfnamefont {Y.}~\bibnamefont
  {Hasegawa}}\ and\ \bibinfo {author} {\bibfnamefont {T.}~\bibnamefont
  {Van~Vu}},\ }\href@noop {} {\bibfield  {journal} {\bibinfo  {journal}
  {Physical Review Letters}\ }\textbf {\bibinfo {volume} {123}},\ \bibinfo
  {pages} {110602} (\bibinfo {year} {2019}{\natexlab{b}})}\BibitemShut
  {NoStop}%
\bibitem [{\citenamefont {Falasco}\ \emph {et~al.}(2020)\citenamefont
  {Falasco}, \citenamefont {Esposito},\ and\ \citenamefont
  {Delvenne}}]{falasco2020unifying}%
  \BibitemOpen
  \bibfield  {author} {\bibinfo {author} {\bibfnamefont {G.}~\bibnamefont
  {Falasco}}, \bibinfo {author} {\bibfnamefont {M.}~\bibnamefont {Esposito}}, \
  and\ \bibinfo {author} {\bibfnamefont {J.-C.}\ \bibnamefont {Delvenne}},\
  }\href@noop {} {\bibfield  {journal} {\bibinfo  {journal} {New Journal of
  Physics}\ }\textbf {\bibinfo {volume} {22}},\ \bibinfo {pages} {053046}
  (\bibinfo {year} {2020})}\BibitemShut {NoStop}%
\bibitem [{\citenamefont {Busiello}\ and\ \citenamefont
  {Pigolotti}(2019)}]{busiello2019hyperaccurate}%
  \BibitemOpen
  \bibfield  {author} {\bibinfo {author} {\bibfnamefont {D.~M.}\ \bibnamefont
  {Busiello}}\ and\ \bibinfo {author} {\bibfnamefont {S.}~\bibnamefont
  {Pigolotti}},\ }\href@noop {} {\bibfield  {journal} {\bibinfo  {journal}
  {Physical Review E}\ }\textbf {\bibinfo {volume} {100}},\ \bibinfo {pages}
  {060102} (\bibinfo {year} {2019})}\BibitemShut {NoStop}%
\bibitem [{\citenamefont {Liepelt}\ and\ \citenamefont
  {Lipowsky}(2007)}]{liepelt1}%
  \BibitemOpen
  \bibfield  {author} {\bibinfo {author} {\bibfnamefont {S.}~\bibnamefont
  {Liepelt}}\ and\ \bibinfo {author} {\bibfnamefont {R.}~\bibnamefont
  {Lipowsky}},\ }\href@noop {} {\bibfield  {journal} {\bibinfo  {journal}
  {Phys. Rev. Lett.}\ }\textbf {\bibinfo {volume} {98}},\ \bibinfo {pages}
  {258102} (\bibinfo {year} {2007})}\BibitemShut {NoStop}%
\bibitem [{\citenamefont {Liepelt}\ and\ \citenamefont
  {Lipowsky}(2009)}]{liepelt2}%
  \BibitemOpen
  \bibfield  {author} {\bibinfo {author} {\bibfnamefont {S.}~\bibnamefont
  {Liepelt}}\ and\ \bibinfo {author} {\bibfnamefont {R.}~\bibnamefont
  {Lipowsky}},\ }\href {\doibase 10.1103/PhysRevE.79.011917} {\bibfield
  {journal} {\bibinfo  {journal} {Phys. Rev. E}\ }\textbf {\bibinfo {volume}
  {79}},\ \bibinfo {pages} {011917} (\bibinfo {year} {2009})}\BibitemShut
  {NoStop}%
\bibitem [{\citenamefont {Gupta}\ and\ \citenamefont
  {Sabhapandit}(2017)}]{gupta2017stochastic}%
  \BibitemOpen
  \bibfield  {author} {\bibinfo {author} {\bibfnamefont {D.}~\bibnamefont
  {Gupta}}\ and\ \bibinfo {author} {\bibfnamefont {S.}~\bibnamefont
  {Sabhapandit}},\ }\href@noop {} {\bibfield  {journal} {\bibinfo  {journal}
  {Physical Review E}\ }\textbf {\bibinfo {volume} {96}},\ \bibinfo {pages}
  {042130} (\bibinfo {year} {2017})}\BibitemShut {NoStop}%
\bibitem [{\citenamefont {Altaner}\ \emph {et~al.}(2015)\citenamefont
  {Altaner}, \citenamefont {Wachtel},\ and\ \citenamefont
  {Vollmer}}]{altaner2015fluctuating}%
  \BibitemOpen
  \bibfield  {author} {\bibinfo {author} {\bibfnamefont {B.}~\bibnamefont
  {Altaner}}, \bibinfo {author} {\bibfnamefont {A.}~\bibnamefont {Wachtel}}, \
  and\ \bibinfo {author} {\bibfnamefont {J.}~\bibnamefont {Vollmer}},\
  }\href@noop {} {\bibfield  {journal} {\bibinfo  {journal} {Physical Review
  E}\ }\textbf {\bibinfo {volume} {92}},\ \bibinfo {pages} {042133} (\bibinfo
  {year} {2015})}\BibitemShut {NoStop}%
\bibitem [{\citenamefont {Ma}\ \emph {et~al.}(2016)\citenamefont {Ma},
  \citenamefont {Hortelao}, \citenamefont {Patino},\ and\ \citenamefont
  {Sanchez}}]{ma2016enzyme}%
  \BibitemOpen
  \bibfield  {author} {\bibinfo {author} {\bibfnamefont {X.}~\bibnamefont
  {Ma}}, \bibinfo {author} {\bibfnamefont {A.~C.}\ \bibnamefont {Hortelao}},
  \bibinfo {author} {\bibfnamefont {T.}~\bibnamefont {Patino}}, \ and\ \bibinfo
  {author} {\bibfnamefont {S.}~\bibnamefont {Sanchez}},\ }\href@noop {}
  {\bibfield  {journal} {\bibinfo  {journal} {ACS nano}\ }\textbf {\bibinfo
  {volume} {10}},\ \bibinfo {pages} {9111} (\bibinfo {year}
  {2016})}\BibitemShut {NoStop}%
\bibitem [{\citenamefont {De~Los~Rios}\ and\ \citenamefont
  {Barducci}(2014)}]{de2014hsp70}%
  \BibitemOpen
  \bibfield  {author} {\bibinfo {author} {\bibfnamefont {P.}~\bibnamefont
  {De~Los~Rios}}\ and\ \bibinfo {author} {\bibfnamefont {A.}~\bibnamefont
  {Barducci}},\ }\href@noop {} {\bibfield  {journal} {\bibinfo  {journal}
  {Elife}\ }\textbf {\bibinfo {volume} {3}},\ \bibinfo {pages} {e02218}
  (\bibinfo {year} {2014})}\BibitemShut {NoStop}%
\bibitem [{\citenamefont {Schnakenberg}(1976)}]{schnakenberg1976network}%
  \BibitemOpen
  \bibfield  {author} {\bibinfo {author} {\bibfnamefont {J.}~\bibnamefont
  {Schnakenberg}},\ }\href@noop {} {\bibfield  {journal} {\bibinfo  {journal}
  {Reviews of Modern physics}\ }\textbf {\bibinfo {volume} {48}},\ \bibinfo
  {pages} {571} (\bibinfo {year} {1976})}\BibitemShut {NoStop}%
\bibitem [{\citenamefont {Vale}\ and\ \citenamefont
  {Milligan}(2000)}]{vale2000way}%
  \BibitemOpen
  \bibfield  {author} {\bibinfo {author} {\bibfnamefont {R.~D.}\ \bibnamefont
  {Vale}}\ and\ \bibinfo {author} {\bibfnamefont {R.~A.}\ \bibnamefont
  {Milligan}},\ }\href@noop {} {\bibfield  {journal} {\bibinfo  {journal}
  {Science}\ }\textbf {\bibinfo {volume} {288}},\ \bibinfo {pages} {88}
  (\bibinfo {year} {2000})}\BibitemShut {NoStop}%
\bibitem [{\citenamefont {Pigolotti}\ \emph {et~al.}(2017)\citenamefont
  {Pigolotti}, \citenamefont {Neri}, \citenamefont {Rold{\'a}n},\ and\
  \citenamefont {J{\"u}licher}}]{pigolotti2017generic}%
  \BibitemOpen
  \bibfield  {author} {\bibinfo {author} {\bibfnamefont {S.}~\bibnamefont
  {Pigolotti}}, \bibinfo {author} {\bibfnamefont {I.}~\bibnamefont {Neri}},
  \bibinfo {author} {\bibfnamefont {{\'E}.}~\bibnamefont {Rold{\'a}n}}, \ and\
  \bibinfo {author} {\bibfnamefont {F.}~\bibnamefont {J{\"u}licher}},\
  }\href@noop {} {\bibfield  {journal} {\bibinfo  {journal} {Physical review
  letters}\ }\textbf {\bibinfo {volume} {119}},\ \bibinfo {pages} {140604}
  (\bibinfo {year} {2017})}\BibitemShut {NoStop}%
\bibitem [{\citenamefont {Pietzonka}\ \emph {et~al.}(2016)\citenamefont
  {Pietzonka}, \citenamefont {Barato},\ and\ \citenamefont
  {Seifert}}]{pietzonka2016universal2}%
  \BibitemOpen
  \bibfield  {author} {\bibinfo {author} {\bibfnamefont {P.}~\bibnamefont
  {Pietzonka}}, \bibinfo {author} {\bibfnamefont {A.~C.}\ \bibnamefont
  {Barato}}, \ and\ \bibinfo {author} {\bibfnamefont {U.}~\bibnamefont
  {Seifert}},\ }\href@noop {} {\bibfield  {journal} {\bibinfo  {journal}
  {Journal of Statistical Mechanics: Theory and Experiment}\ }\textbf {\bibinfo
  {volume} {2016}},\ \bibinfo {pages} {124004} (\bibinfo {year}
  {2016})}\BibitemShut {NoStop}%
\bibitem [{\citenamefont {Dass}\ \emph {et~al.}(2021)\citenamefont {Dass},
  \citenamefont {Georgelin}, \citenamefont {Westall}, \citenamefont {Foucher},
  \citenamefont {De~Los~Rios}, \citenamefont {Busiello}, \citenamefont
  {Liang},\ and\ \citenamefont {Piazza}}]{dass2021equilibrium}%
  \BibitemOpen
  \bibfield  {author} {\bibinfo {author} {\bibfnamefont {A.~V.}\ \bibnamefont
  {Dass}}, \bibinfo {author} {\bibfnamefont {T.}~\bibnamefont {Georgelin}},
  \bibinfo {author} {\bibfnamefont {F.}~\bibnamefont {Westall}}, \bibinfo
  {author} {\bibfnamefont {F.}~\bibnamefont {Foucher}}, \bibinfo {author}
  {\bibfnamefont {P.}~\bibnamefont {De~Los~Rios}}, \bibinfo {author}
  {\bibfnamefont {D.~M.}\ \bibnamefont {Busiello}}, \bibinfo {author}
  {\bibfnamefont {S.}~\bibnamefont {Liang}}, \ and\ \bibinfo {author}
  {\bibfnamefont {F.}~\bibnamefont {Piazza}},\ }\href@noop {} {\bibfield
  {journal} {\bibinfo  {journal} {Nature Communications}\ }\textbf {\bibinfo
  {volume} {12}},\ \bibinfo {pages} {1} (\bibinfo {year} {2021})}\BibitemShut
  {NoStop}%
\bibitem [{\citenamefont {Ito}(2018)}]{ito2018stochastic}%
  \BibitemOpen
  \bibfield  {author} {\bibinfo {author} {\bibfnamefont {S.}~\bibnamefont
  {Ito}},\ }\href@noop {} {\bibfield  {journal} {\bibinfo  {journal} {Physical
  review letters}\ }\textbf {\bibinfo {volume} {121}},\ \bibinfo {pages}
  {030605} (\bibinfo {year} {2018})}\BibitemShut {NoStop}%
\bibitem [{\citenamefont {Nicoletti}\ and\ \citenamefont
  {Busiello}(2021)}]{nicoletti2021mutual}%
  \BibitemOpen
  \bibfield  {author} {\bibinfo {author} {\bibfnamefont {G.}~\bibnamefont
  {Nicoletti}}\ and\ \bibinfo {author} {\bibfnamefont {D.~M.}\ \bibnamefont
  {Busiello}},\ }\href@noop {} {\bibfield  {journal} {\bibinfo  {journal}
  {Physical review letters}\ }\textbf {\bibinfo {volume} {127}},\ \bibinfo
  {pages} {228301} (\bibinfo {year} {2021})}\BibitemShut {NoStop}%
\end{thebibliography}

\end{document}